\documentclass[lettersize,journal]{IEEEtran}
\usepackage{amsmath,amsfonts}
\usepackage{algorithmic}
\usepackage{algorithm}
\usepackage{array}
\usepackage[caption=false,font=normalsize,labelfont=sf,textfont=sf]{subfig}
\usepackage{textcomp}
\usepackage{stfloats}
\usepackage{url}
\usepackage{verbatim}
\usepackage{graphicx}
\usepackage{cite}
\usepackage{mathptmx}  
\usepackage{multirow}
\usepackage{makecell}
\usepackage{booktabs}
\usepackage{bm}
\usepackage{xcolor}
\definecolor{green}{HTML}{00CC66}
\makeatletter
\renewcommand{\@IEEEsectpunct}{\ \,} 
\makeatother

\begin{document}

\title{Multidimensional Physiology-Inspired Enhanced Vital Sign Monitoring Using MIMO mmWave Bio-radar}

\author{Heyao Zhu, Yimeng Zhao, Zirui Zhang, Huansheng Yi, Chenbin Gao, Canhua Xu, Jianqi Wang and Fugui Qi, ~\IEEEmembership{Member,~IEEE}
\thanks{Manuscript received XXX, XXXX; revised XXX, XXXX. This work was supported in part by the National Natural Science Foundation of China under Grant 62201578; in part by Young Talent Fund of Association for Science and Technology in Shaanxi, China (20230144); and in part by the Key Research and Development Plan of Shaanxi Province under Grant 2024SF-YBXM-658. \textit{(Corresponding author: Fugui Qi)}}
\thanks{Heyao Zhu, Zirui Zhang, Chenbin Gao are with the School of Basic Medicine, Fourth Military Medical University, Xi'An 710032, China.(e-mail: 2658069817@qq.com, 1075768766@qq.com, 3402995528@qq.com)}
\thanks{Yimeng Zhao is with the School of Mechanical and Electrical Engineering, Chengdu University of Technology, Chengdu 610059, China (e-mail: yimeng\_zhao@stu.cdut.edu.cn)}
\thanks{Huansheng Yi, Canhua Xu, Jianqi Wang are with the School of Biomedical Engineering, Fourth Military Medical University, Xi'an 710032, China (e-mail: yihsbme@outlook.com, canhuaxu@fmmu.edu.cn, wangjq@fmmu.edu.cn;)}
\thanks{Fugui Qi is with the School of Biomedical Engineering, Fourth Military Medical University, Xi'An 710032, China and also with Military Medical Innovation Center, Xi'an 710032, China (e-mail: Qifgbme@outlook.com)}}

\markboth{Journal of \LaTeX\ Class Files,~Vol.~14, No.~8, August~2021}%
{Shell \MakeLowercase{\textit{et al.}}: A Sample Article Using IEEEtran.cls for IEEE Journals}

\IEEEpubid{0000--0000/00\$00.00~\copyright~2021 IEEE}


\maketitle

\begin{abstract}
With the intensification of population aging and  increasing burden of chronic diseases, the demand for  vital signs monitoring is becoming increasingly urgent. A key challenge facing current non-contact detection technologies using millimeter wave (mmWave) radar is the low efficiency of multi-channel signal fusion in array radar systems based on equal weighting. To address this challenge, this paper proposes a vital sign enhancement detection method for multiple input and multiple output (MIMO) bio-radar, driven by multidimensional physiological characteristics, which overcomes traditional limitations through a two-stage fusion strategy. Stage 1: Enhanced Vital Sign Detection Using Single-Channel Signals Based on Physiological Characteristics. First, a chest wall multi-scattering point model is constructed. For single channel time-distance two-dimensional echo signals, effective range bins are selected based on the respiratory/cardiac physiological frequency band energy ratio, and the signal-to-noise ratio (SNR) of respiration/heart signals is enhanced using phase-aligned maximal ratio combining (MRC). Stage 2: Multi-Channel Fusion Based on Organ Radiation Spatial Distribution Characteristics. The spatial radiation characteristics of cardiopulmonary organs are introduced for the first time as the theoretical foundation for SNR-based channel screening, channel attribute identification, and multi-channel weighted fusion. Then, we propose a template matching method to extract respiratory rate (RR) and heart rate (HR) by adopting physical models of respiration and cardiac activities. The experimental results demonstrate the existence of the spatial distribution characteristics of organ radiation. In addition, we analyzed the impact of distance and state on the algorithm from these two aspects. The results show that our method outperforms the reference methods and exhibits strong robustness when scenarios change.

\end{abstract}


\section{Introduction}
\IEEEPARstart{W}{ith} the aggravation of population aging and the aggravation of chronic disease burden, vital signs monitoring technology has become an important research direction in the field of medical health. By the end of 2023, the population aged 60 and above accounted for 21\%. According to the World Population Prospects 2022, China's elderly population (aged 60 and above) is projected to rise significantly and peak around 2050 at approximately 580 million\cite{un_wpp2022}. At the same time, Cardiovascular and cerebrovascular diseases (such as stroke and ischemic heart disease) and chronic respiratory diseases rank among the leading causes of mortality \cite{2b}. This trend significantly increases the need for vital signs monitoring. Traditional contact devices (such as ECG) have high accuracy, but there are obvious drawbacks. One significant drawback is discomfort during wear. This discomfort may lead users to be reluctant to wear it or to remove it midway, thereby affecting the continuity of vital signs monitoring. Another drawback is that wearable devices are prone to causing allergic reactions. For example, smart watches may cause contact dermatitis due to acrylate allergy \cite{4}, and some elderly users are prone to skin itching after wearing ECG monitoring devices \cite{3}. Under this background, non-contact detection technology based on millimeter wave (mmWave) radar, due its advantages such as high user comfort and continuous monitoring capability. 

Although FMCW radar technology \cite{7} has made significant progress in the field of non-contact vital signs monitoring\cite{8}, existing research still faces key challenges in terms of detection accuracy and robustness, mainly reflected in three aspects: model assumptions, signal fusion strategies and dynamic scene adaptability. \textit{1)} Single point model assumption error. \IEEEpubidadjcol Current mainstream methods generally model the human chest wall as a single scattering point\cite{9,10,11,12,13,14,15}, which ignores its actual spatial geometric distribution (the thoracic thickness of an adult is about 20 to 30 cm). This simplified model becomes particularly ineffective at close ranges ($<$1 m) due to significant biases in the selection of effective range bins. As noted in\cite{9}, this bias introduces up to 20\% or more heart rate (HR) estimation error. In addition, the model assumes that the subject must face the radar directly, making it sensitive to changes in posture and orientation. It is also vulnerable to strong reflective clutter (such as clothing wrinkles and furniture reflections), and exhibits poor robustness under varying detection conditions, such as changes in distance or angle. \textit{2)} Low signal fusion efficiency. Traditional processing methods (such as incoherent superposition or mean fusion) ignore the phase differences of respiratory and heartbeat signals (0-200 ms delay), resulting in signal distortion, especially when the signal-to-noise ratio (SNR) decreases by $>$5 dB in the presence of random body motion (RBM). \textit{3)} Insufficient adaptability to dynamic scenes. Existing algorithms are sensitive to dynamic scenarios such as subject position shifts or slight body movements (e.g., head turning or leg shaking), which significantly affect the stability and accuracy of HR detection, thereby limiting their generalization capability in real-world applications.

Based on the advantages of multiple input and multiple output (MIMO) radar multi-channel detection, some previous studies initially explored the use of multi-channel fusion enhancement detection, but its fusion strategy has obvious limitations. For example, Reference\cite{17} processes the spectral output of each channel using incoherent superposition (FFT amplitude superposition). Such methods rely solely on the amplitude information of the signal (power spectrum), completely discarding the phase information, which is rich in target displacement details and spatial correlations. Moreover, the indiscriminate, ``uniform-weight'' summation or averaging of echoes from different spatial locations leads to signal distortion. Another fusion idea, such as Maximal Ratio Combining (MRC) adopted in \cite{19}, is theoretically optimal, but it fuses all 192 channels (including inferior and even invalid channels). This fails to distinguish the differences in the distribution of cardiopulmonary organ activity in the thoracic and abdominal surface spaces, possibly introducing noise or low-quality channel information into the fusion results, limiting its potential gain. The study in \cite{20} utilizes the range resolution of FMCW by selecting effective range units based on the frequency characteristics of cardiopulmonary activity and performing phase-coherent fusion, similar to using a single beam to irradiate specific areas of the chest and abdomen. However, these methods mainly rely on preset templates for range bin selection, and fail to fully utilize all the information provided by MIMO radar in the whole thoraco-abdominal region in spatial dimension (angle/channel). In summary, existing fusion methods are inefficient in utilizing spatial information, resulting in increased errors, susceptibility to interference (especially RBM), and insufficient robustness to changes in detection conditions.

The inherent unique attributes of cardiopulmonary activity, i.e., the physiological characteristics (frequency, energy) and spatial distribution characteristics of body surface micromotion caused by respiration and heartbeat, provide an opportunity to overcome the above challenges: \textit{1)} Physiological characteristics. Heartbeat signal frequency is higher (0.8-2 Hz), energy is usually much lower than respiration and its harmonics. Respiration signal frequency is lower (0.1-0.8 Hz), energy is higher and there are obvious harmonic components. This feature helps in the preliminary differentiation of cardiac and respiratory signals. \textit{2)} Spatial distribution characteristics of organ radiation. Significant body surface micromotion caused by cardiac activity (apical beat) is mainly confined to a specific region of the left chest (the precordial region). Respiratory activity varies according to the breathing pattern (thoracic or abdominal respiration). Micromotion caused by thoracic breathing is mainly distributed in the chest, while micromotion caused by abdominal breathing is mainly distributed in the abdomen. Even for the same breath, the respiratory rhythm information captured from different regions of the chest and abdomen must differ in both amplitude and phase. Therefore, by considering physiological characteristics, incorporating the spatial correspondence of surface micromotions generated by different organs (heart and lungs) in distinct regions of the chest and abdomen, and carrying out targeted signal selection and fusion, it is expected to significantly enhance the target vital sign signal and suppress noise and interference.

To this end, leveraging the multi-channel sensing advantages provided by MIMO FMCW radar and considering the physiological characteristics of cardiopulmonary activity as well as the spatial distribution characteristics of organ radiation, this paper proposes a novel high-precision and robust vital signs enhancement detection method based on MIMO bio-radar. The core innovations and contributions of this study are as follows:

\begin{enumerate}
  \item{Effective range bins selection and fusion driven by physiological characteristics based on multi-scattering point model (single channel) \cite{21}. \textit{a)} For the echo signals obtained from single-channel detection of micro-movements in the human chest and abdomen, we first break through the limitation of single scattering point model and establish a multi-scattering point model for single beam irradiating the partial chest wall. \textit{b)} On this basis, multiple effective range bins are intelligently selected from the single-channel time-distance two-dimensional echoes, by incorporating the physiological characteristics of cardiopulmonary activity. \textit{c)} Multiple effective range bins coherent fusion. Maximum ratio combining (MRC) is used to perform phase alignment weighted fusion on multiple effective range bin signals to improve SNR.
  }
  \item{Channel identification and fusion driven by spatial distribution characteristics of organ radiation (multichannel). \textit{a)} SNR-dominated channel screening. For the multi-channel echoes obtained from MIMO FMCW coverage detection of chest and abdominal micromotion, a cardiopulmonary activity information quality evaluation and channel screening is performed based on respiratory and heartbeat SNR parameters. \textit{b)} Channel attribute identification. Channels are classified into respiration-dominant, heartbeat-dominant, and mixed types based on the ratio between the respiration signal SNR and the heartbeat signal SNR, with corresponding signal components selectively retained. \textit{c)} Weighted fusion of different organ regions (channels). MRC fusion is performed on signals of the same category with SNR as weight.\cite{22}}
  \item{Based on a series of experiments conducted under varying relative illumination coverage areas and different detection conditions, the existence of the spatial distribution characteristics of organ radiation and the effectiveness of the proposed method are verified.}
\end{enumerate}

\indent The rest of this paper is structured as follows: Section II introduces the system architecture and multi-scattering point signal model. Section III introduces the proposed method in detail. Section IV shows the experimental design and data collection process. Section V shows the results and discussion. Section VI concludes the paper.

\section{SYSTEM ARCHITECTURE AND MULTI-SCATTERING POINT SIGNAL MODEL}

\begin{figure*}[htbp]
    \centering
    \includegraphics[width=\textwidth]{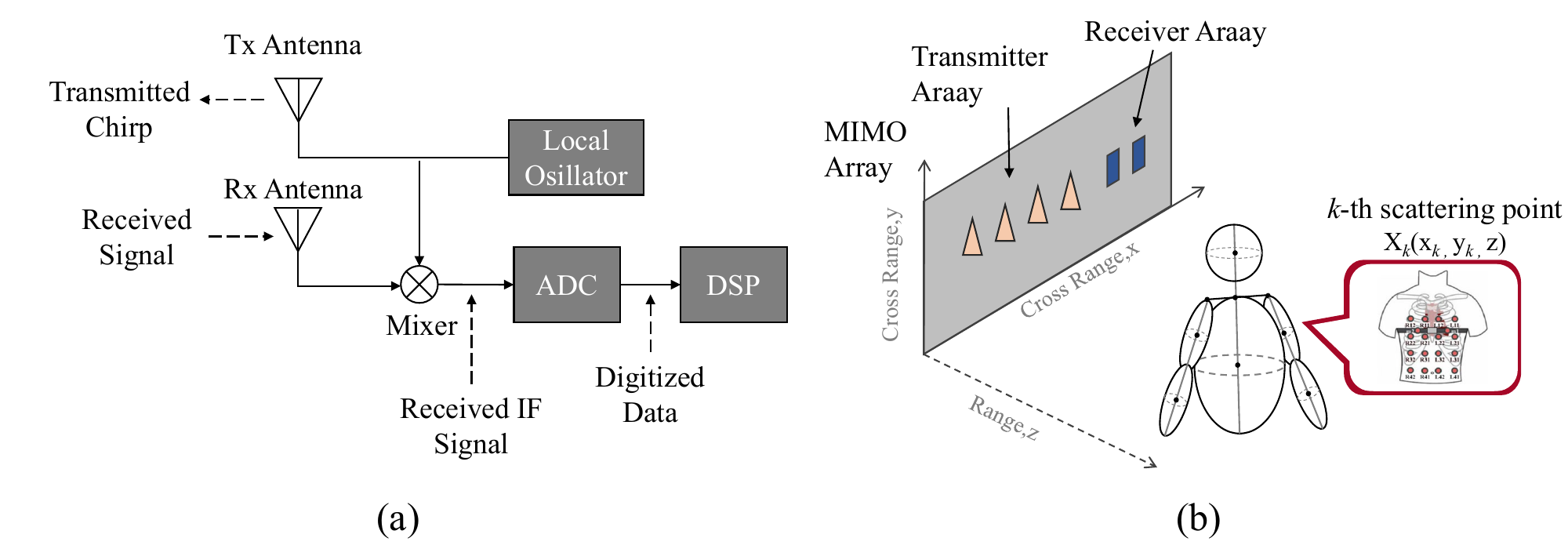} 
    \caption{
  (a) Radar structure block diagram.
  (b) Schematic diagram of radar area array detecting multi-point vibration of chest and abdomen.}

  \label{fig1}
  \end{figure*}

For a MIMO FMCW radar system (Fig. 1 (a)), $M$ transmit antennas (TX) and $N$ receive antennas (RX) form $\textit{M} \times \textit{N}$ channels through time division multiplexing (TDM-MIMO) techniques. Radar transmits a linear FMCW signal:

\begin{equation}
  \mathrm{s}(t)=A_t \cos \left(2 \pi f_c t+\pi S t^2\right), 0 \leq t \leq T_c
\end{equation}
where carrier frequency $f_c$ = 77 GHz, FM slope $S = B/T_{c} = 66$ GHz/s, bandwidth $B = 4$ GHz, $T_c$ are pulse widths.

\subsection{Multi-scattering point model and single coupling}
As shown in Fig. 1 (b), the human chest wall can be modeled as $K$ scattering points. The $k$-th scattering point is located at a distance $R_k(t)$, and its displacement $d_k(t)$ is modulated by the respiratory motion $d_r(t)$ and the heartbeat motion $d_h(t)$:

\begin{equation}
  R_k(t)=R_{k 0}+d_k(t)=R_{k 0}+\alpha_k d_r(t)+\beta_k d_h(t)
\end{equation}
where $\alpha_k$, $\beta_k$ are position-dependent modulation coefficients (e.g. apical region $\beta _{k}\gg \alpha _{k}  $).

\subsection{Multi-channel received signal model}
Through the TDM-MIMO system, the virtual channel reception signal formed by the $m$-th transmitting antenna and the $n$th receiving antenna is:
\begin{equation}
  s_{m n}(t)=\sum_{k=1}^K C_{m n k} \exp \left\{j\left[\frac{4 \pi}{\lambda} R_k(t)+\phi_{m n k}+\Delta \phi(t)\right]\right\}
\end{equation}
Where $C_{mnk}$ is the channel gain (including antenna pattern and path loss), $\phi _{mnk}$ is the fixed phase offset (determined by antenna position and target angle), and $\Delta \phi (t)$ is the phase noise term.

\subsection{IF signal decoupled from displacement}
The IF phase after mixing contains target displacement information:
\begin{equation}
  \phi_{m n k}^{I F}(t)=\frac{4 \pi}{\lambda} d_k(t)+\phi_{m n k}
\end{equation}

The signals of scattering points are separated by Range-Doppler processing (Range-FFT and Doppler-FFT), and spatial resolution is realized by combining DOA estimation. Finally, the multi-point displacement $d_k(t)$ is extracted. 

\section{Method}
\begin{figure*}[htbp]
    \centering
    \includegraphics[width=\textwidth]{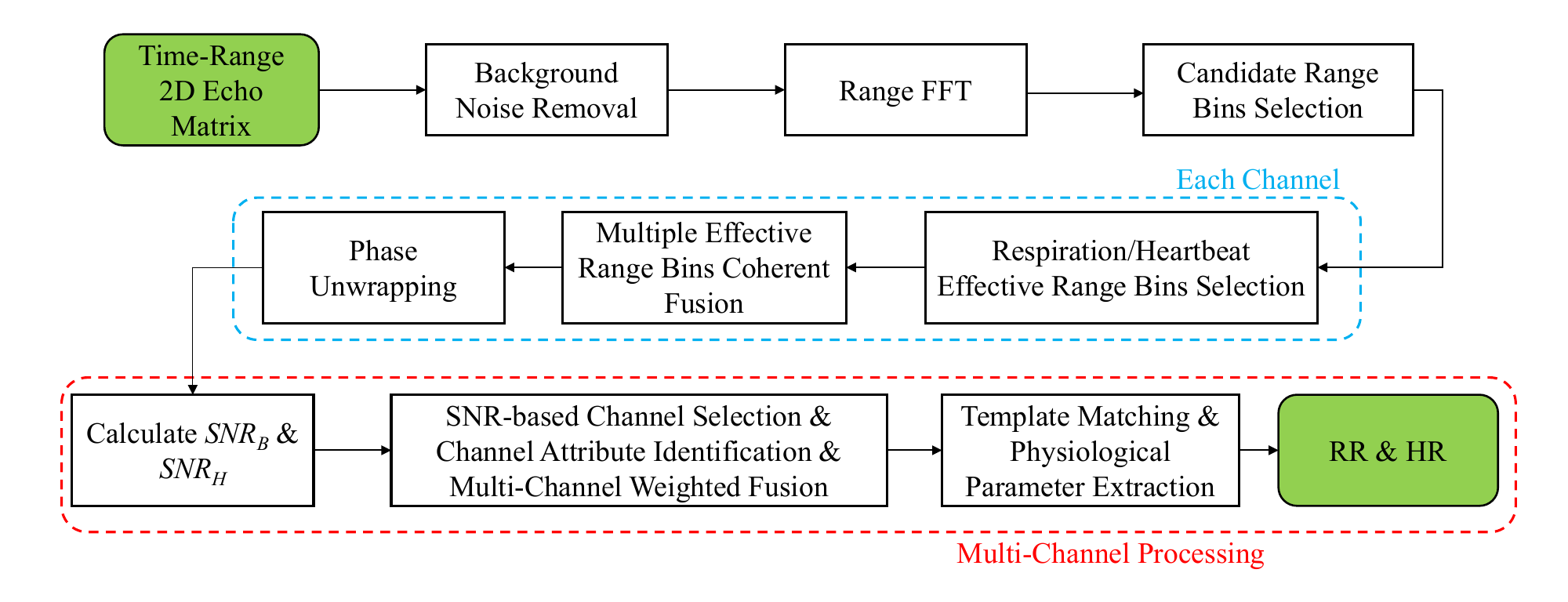} 
    \caption{Signal processing flow.}
  \label{fig2}
  \end{figure*}
  \begin{figure}[htbp]
    \centering
    \includegraphics[width=0.5\textwidth]{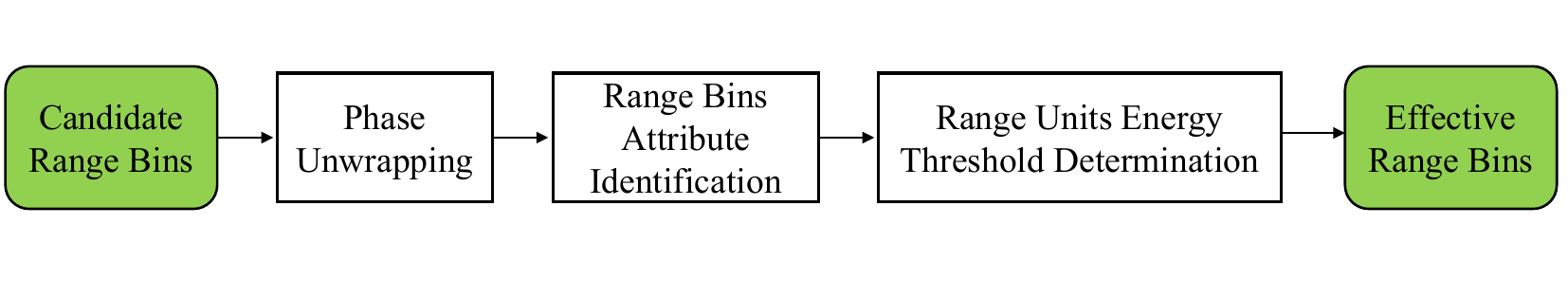} 
    \caption{Flow chart of effective range bins selection.}

  \label{fig3}
  \end{figure}
According to the signal processing framework shown in Fig. 2, the proposed method systematically integrates the concept of multi-scatterer modeling with the spatial distribution characteristics of organ radiation. It includes three key components: single-channel signal enhancement, multi-channel fusion enhancement detection incorporating the spatial distribution characteristics of organ radiation, and physiological parameter extraction. This section will introduce each component in detail.
\subsection{Preprocessing and effective range bins selection}
\subsubsection{Preprocessing.} After acquiring the echo signals, radar data from 8 channels are obtained through data parsing. The single-channel signal processing flow is introduced using any one of the channels as a representative. The single-channel echo signal undergoes DC offset removal, sidelobe suppression using a Blackman window, and fast Fourier transform (FFT) along the range dimension to generate a time-distance matrix.

\subsubsection{Effective distance bins selection.} First, the CA-CFAR method is applied to traverse each range bin and select those where the target signal is likely present as candidate bins. Then, effective range bins are determined from the candidates using an energy ratio-based method \cite{13}. The steps of the energy ratio method are as follows:

\textbf{Frequency band division.} The frequency of the respiration signal ranges from 0.1 to 0.8 Hz, and that of the heartbeat signal ranges from 0.8 to 2 Hz. The energy peak is checked to determine whether it falls within the respiratory or heartbeat frequency band. If not, the corresponding bin is excluded; if it does, the process proceeds to the next step. At this point, the distance unit dominated by the respiratory signal and the distance unit dominated by the heartbeat signal are divided.

\textbf{Energy threshold determination.} Respectively calculate the energy within and outside the corresponding frequency bands for respiration- and heartbeat-dominant range bins. If ${E_{in}}/{E_{out}}>5$, the range bin is defined as an effective range bin. The effective range bins selection flow chart is shown in Fig. 3.

\subsection{Coherent fusion of multiple range bin signals}
Previous studies have shown that the movements of multiple regions on the human chest wall exhibit correlation and delay characteristics \cite{23}, and that coherent fusion \cite{20} can improve signal SNR. To this end, this study performs coherent fusion of multiple effective range bin echo signals obtained from the single channel. The algorithm implementation mainly includes the following four step.
\subsubsection{Signal modeling.} Different range bins receive physiological signals arriving via different propagation paths. Each path has an independent propagation coefficient, which is influenced by factors such as chest wall thickness, surface reflectivity, path length, and environmental noise. Therefore, the signal received by each range bin can be expressed as:
\begin{equation}
  y_{i} =h_{i}\ast s+n
\end{equation}
where $s$ represents motion at the chest wall, $h_i$ is the transfer coefficient, and $n$ is noise. The signal matrix of the effective range bins is represented as $Y=hs+N$, and the dimension of $Y$ is $I \times M$, where $I$ is the number of effective range bins and $M$ is the number of slow time points.

\subsubsection{Phase delay estimation.} The cross-correlation function is used to calculate the mutual correlation of phase signals from all effective range bins. The bin with the highest sum of coherence with all other bins is selected as the reference bin. Using this reference bin as a baseline, the time delays $\phi _{i}$ of the other bins are calculated.

\subsubsection{Transfer coefficient estimation.} Construct a phase delay vector $e=[e^{j\phi _{1} },\dots ,e^{j\phi _{I} } ]^{T} $, estimate chest wall motion by minimizing noise $\hat{s}$:
\begin{equation}
  \hat{s} =e^{H} Y
\end{equation}
and then infer the transmission coefficients $h$ using $\hat{s}$.

\subsubsection{Maximum ratio combining weighted fusion.} The weight of each effective range unit depends on the size of its transfer coefficient ($\omega _{i}\propto \left | h_{i}  \right |  $). After calculating the weight, the signal of each unit is multiplied by its corresponding weight value respectively, and then the results are summed to complete the coherent fusion of multiple effective range bin signals. As a result, the fused physiological signal of one channel is obtained. By repeating the above steps, the physiological signals of all $M \times N$ channels can be obtained.

\begin{figure}[htbp]
    \centering
    \includegraphics[width=0.5\textwidth]{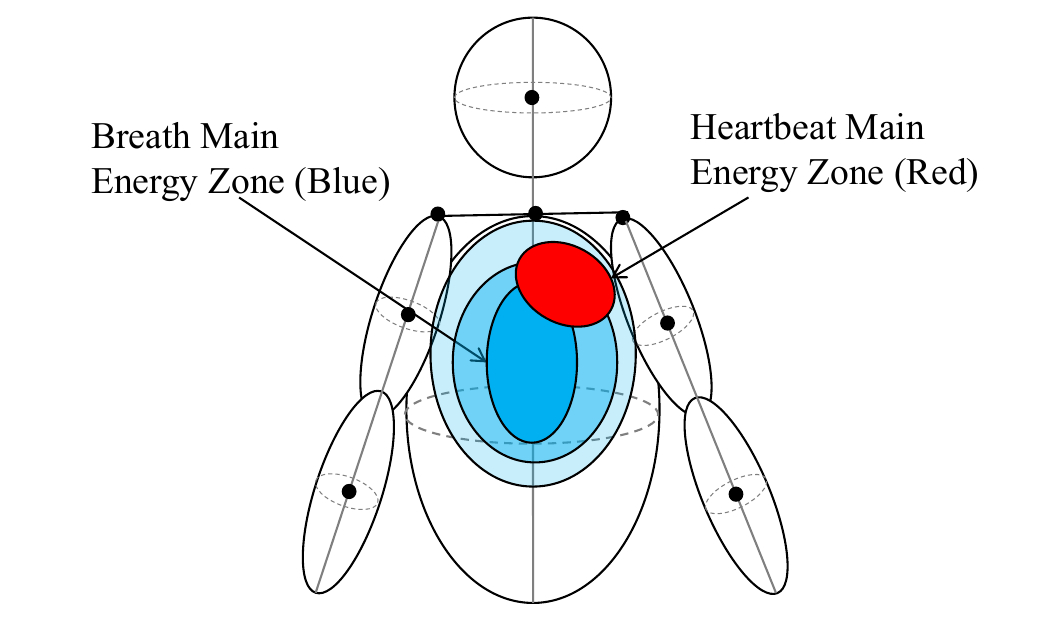} 
    \caption{Schematic diagram of spatial distribution characteristics of organs radiation.}

  \label{fig4}
  \end{figure}

\subsection{Multi-channel fusion enhancement detection based on spatial distribution characteristics of organ radiation} 
The above process only involves the processing of signals detected by a single channel of the MIMO radar from specific regions of the chest and abdomen, without fully considering the spatial distribution characteristics of the heart and lungs, thereby limiting detection accuracy. Therefore, in order to fully leverage the advantages of MIMO radar multi-channel detection and further enhance detection performance by incorporating the spatial distribution characteristics of organ radiation, we propose an innovative method for channel identification and fusion driven by the spatial distribution characteristics of organ radiation.

We first elaborate in detail on the modeling and analysis of the spatial distribution characteristics of organ radiation, providing a theoretical basis for multi-channel fusion algorithms.

\textbf{Organ radiation spatial distribution modeling.} According to the beam irradiation analysis of mmWave array radar detecting chest and abdomen micromotion, the physiological activities of cardiopulmonary organs will form corresponding radiation area distribution on the body surface (as shown in Fig. 4). \textit{1)} Apical pulsation area (red area). The apical pulsation area is located within a spatial range of 5-8 cm, with a diameter between the 4th and 5th ribs along the left midclavicular line. Its movement characteristics include periodic pulsation at 0.8-2 Hz, with an average displacement of 0.6 ± 0.2 mm. \textit{2)} Respiratory motor zone (blue area). The respiratory motor zone spans a longitudinal belt from the sternal angle to the xiphoid process, with a width of 12-15 cm. Its movement characteristics involve rhythmic fluctuations between 0.1-0.8 Hz, with a displacement amplitude of 2.5 ± 0.8 mm.

\textbf{Analysis of spatial distribution characteristics of organ radiation.} Different physiological signals exhibit spatial distribution differences. The main energy zone of cardiac pulsation is concentrated in the area between the 4th and 5th ribs on the left chest (the apical pulsation point). The main energy zone of respiratory motion is distributed along a longitudinal belt area from the sternal angle to the xiphoid process.

Based on the above model and characteristic analysis, the specific steps of the innovative algorithm proposed in this paper are as follows:

\subsubsection{Channel screening based on SNR.} The goal is to identify and eliminate inferior channels based on SNR. First, the SNR of respiration and heartbeat are calculated, i.e. $SNR_{B}=\left \{ SNR_{B1} , SNR_{B2}, \dots , SNR_{BN} \right \}  $ and $SNR_{H}=\left \{ SNR_{H1} , SNR_{H2}, \dots , SNR_{HN} \right \} $. $SNR_B$ and $SNR_H$ are sorted in descending order respectively. The significant slope mutation points are determined using the first-order difference, and the mean values $\mu _{B} $ and $\mu _{H}$ of subset before abrupt points are calculated.  Channels $i$ that satisfy both conditions of $SNR_{B,i} < 0.8\mu _{B} $ and $SNR_{H,i} < 0.8\mu _{H}$ are eliminated.

\subsubsection{Channel attribute identification and weighted fusion.} The goal is to suppress motion interference from ineffective regions and reduce mutual interference between the heart and lungs, focusing on the effective vital sign information corresponding to the heart/lung regions. If respiration (heartbeat) is the dominant component, the respiration (heartbeat) signal is retained for subsequent detection. Specifically, if the $SNR_B/SNR_H > $ 3, it is considered to be a respiratory dominant channel (a high threshold ensures that only channels with significant respiratory signal dominance are selected to avoid interference from heartbeat signals or noise). If the $SNR_H/SNR_B < $ 0.35, it is considered to be the heartbeat dominant channel (heartbeat signal amplitude is small and easily modulated by respiration, and a low threshold is required to capture valid heartbeat components). In other cases, it is considered a mixed component channel, which may contain coupled information from respiration and heartbeat, and retaining both enhances signal integrity at fusion. Finally, respiratory and heartbeat signals are fused separately with SNR as weights.

\subsection{Template matching and physiological parameter extraction}


Based on the fused signal obtained from the previous step, a template matching method is used to extract physiological parameters. This method constructs motion models for respiration and heartbeat to generate corresponding template signals. It then iterates through the parameter value range to find the optimal shaping parameters that maximize the cross-correlation with the actual radar signal, thereby achieving high-precision extraction of the respiration rate (RR) and HR.

According to previous studies\cite{24}, template signals can be generated based on RC circuit models of respiration and relaxation oscillation systems of heartbeat, with the specific models as follows:

\textbf{Respiratory model.}
Breathing is a periodic activity involving airflow in and out of the lungs. Contraction and relaxation of the diaphragm cause changes in pleural pressure and control airflow into and out of the lungs. The chest wall movement, which is proportional to the change in lung volume, can be represented by an RC circuit to model the process of respiration. The solution of lung volume is:
\begin{equation}
  V(\mathrm{t})=\left\{\begin{array}{l}
\frac{\tau R S}{R_{r s}}\left[A_1 t^2+A_2 t+A_3\left(1-e^{\frac{t}{\tau R S}}\right)\right]+V_0 e^{\frac{t}{\tau R S}}, 0 \leq t \leq t_1 \\
\frac{P\left(t_1\right)}{R_{r s}\left(\frac{1}{\tau R S}-\frac{1}{\tau}\right)}\left(e^{\frac{t-t_1}{\tau}}-e^{\frac{t-t_1}{\tau R S}}\right)+V\left(t_1\right) e^{\frac{t-t_1}{\tau R S}}, t_1 \leq t \leq t_1+t_2
\end{array}\right.
\end{equation}

\begin{table}[htbp]
  \centering
  \caption{Main parameters of radar}
    \begin{tabular}{cc}
    \hline
    \specialrule{0em}{0.5pt}{0.5pt}
    \hline
    \textbf{Parameter Name} & \textbf{Parameter Value} \\
    \midrule
    Number of transmit antennas & 2 \\
    Number of receive antennas & 4 \\
    Chirp slope & 66 GHz/s \\
    Start frequency & 77 GHz \\
    End frequency & 81 GHz \\
    Bandwidth & 4 GHz \\
    Number of range bins & 256 \\
    Slow-time sampling rate & 20 Hz \\
    \hline
    \specialrule{0em}{0.5pt}{0.5pt}
    \hline
    \end{tabular}%
  \label{tab1}%
\end{table}%

\textbf{Heartbeat model.} Within the thorax, the heart is located at the top of the diaphragm, with the apex near the front surface of the thorax. When the heart beats, contraction and relaxation change the distance between the apex and the chest wall surface, causing chest wall movement. The rhythm of the heart is generated by the sinoatrial node and conducted from the atria to the ventricles along the bundle of His and Purkinje fibers. The rhythm of the sinoatrial node behaves as a relaxed oscillatory system, i.e. the natural frequency of the heart adapts to the drive of external frequencies without changing its amplitude characteristics. The system is defined as:

\begin{equation}
  \frac{\partial^2 V}{\partial t^2}-\alpha\left(1-V^2\right) \frac{\partial V}{\partial t}+V=0
\end{equation}

Matching the fusion signal with template shaping parameters, the template signal with maximum cross-correlation with the fusion signal is searched, and RR and HR are extracted.

\section{EXPERIMENTAL SETUP AND DATA COLLECTION}
\subsection{Experiment platform configuration}
The AWR1642 radar is mounted on a height-adjustable support with a horizontal coverage angle of 120° and a vertical coverage angle of 22.2°. The acquired original IF signal is sent to the host computer through DCA1000EVM acquisition card. mmWave Studio is used to start, configure and control radar boards. In the experiment, the signal measured by respiratory heartbeat bandage is taken as reference truth value. The signal processing flow is implemented in MATLAB R2023a, taking the raw IF signal as input and outputting the vital sign signal waveform as well as the estimated RR and HR. Main parameters of radar are shown in Table $\mathrm{I}$.


\subsection{Experiment design}
To assess the performance of our proposed method, eight healthy subjects (aged 19-30 years, with an equal number of males and females) were recruited and tested in a laboratory environment (6.5 m $\times$ 5 m) while wearing daily clothing. The Ethic Committee of the First Affiliated Hospital of the Fourth Military Medical University approved the study (No. KY20253449-1). The informed consent of all subjects was obtained prior to their participation.

\begin{figure}[htbp]
    \centering
    \includegraphics[width=0.5\textwidth]{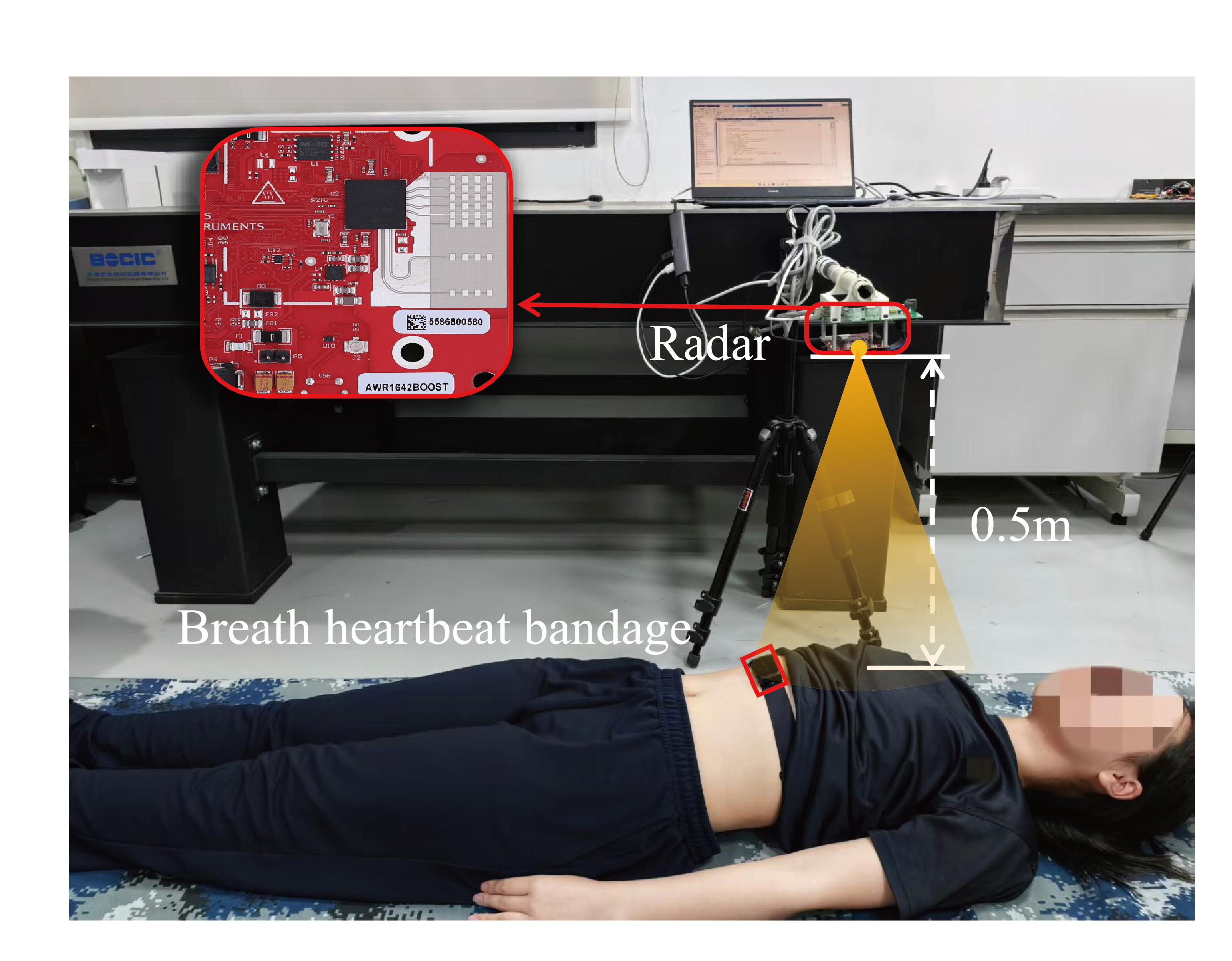} 
    \caption{Experimental scenario.}

  \label{fig5}
  \end{figure}
\subsubsection{Experimental design for validating the spatial distribution characteristics of organ radiation}
\
\newline
\indent In order to verify the existence of spatial distribution characteristics of organ radiation, the following validation experiment was designed: 4 subjects were recruited and lay still at a distance of 0.5 m from the radar, the experimental scenario is shown in Fig. 5. During the experiment, the radar position was fixed, and the chest and abdomen region of the subject was detected. The subjects shifted their position to ensure that the radar's effective irradiation area sequentially covered the right half area, the entire area and the left half area. The data collection time for each set of experiments is 15s. Finally, A total of 12 sets of experimental data (4 subjects $\times$ 3 irradiation areas) were collected for the subsequent analysis and validation of the spatial distribution characteristics of organ radiation.

\subsubsection{A series of experiments under different detection conditions}
\
\newline
\indent Each subject performed four states: sitting (SI), standing (ST), leg shaking (LS), and head turning (HT) at three distances (0.3 m, 1 m, and 2 m), That is, the experimental scenarios consists of a combination of distance variables $r$ = [0.3 m, 1 m, 2 m] and state variables $s$ = [SI, ST, LS, HT]. Each state was repeated three times to reduce random errors, with each trial lasting 15 seconds. The total amount of data is 288 (8 subjects $\times$ 3 distances $\times$ 4 states $\times$ 3 repeats). Data measured by the respiration and heartbeat bands indicate that the subjects' RR varies between 11 and 27 bpm, while HR ranges from 65 to 104 bpm.

\subsection{Reference methods and performance evaluation indicators}

To verify the effectiveness of the proposed method, two typical MIMO radar multi-channel fusion methods are used as references. Reference method 1 (Ref1) \cite{12},  which is a multi-channel fusion enhancement detection method based on signal correlation, adopts maximum ratio combining (MRC) for weighted fusion of multi-channel signals based on the signal correlation of all channels, with only one effective distance unit selected per channel. It then uses continuous wavelet transform (CWT) to extract RR and HR. But to control for variables and demonstrate the superiority of our proposed method, we use the same physiological parameter extraction method (template matching). Ref1 does not perform single-channel signal fusion. Therefore, we further propose an improved reference method 2 (Ref2), which is a multi-channel fusion enhanced detection method based on multiple range bins and signal correlation. This method combines the multi-scattering point model and fuses the multiple range bin signals weighted by phase coherence to obtain a single-channel processed signal. The subsequent processing is consistent with that of Ref1.

\textbf{Evaluate metrics.} The objective of this paper is to extract vital sign information from radar signals. We select the most commonly used metric to quantify the accuracy of the proposed algorithm: error rate, calculated as equation (9), where $R_e$ is the RR/HR estimated by the algorithm, and $R_t$ is the true value measured by the respiratory heartbeat bandage.

\begin{equation}
  Error=\frac{\left | R_{e}-R_{t}   \right | }{R_{t} }  \times 100\%
\end{equation}

\begin{table*}[htbp]
  \centering
  \caption{$SNR_B$ and  $SNR_H$ average values in the right half area.}
    \begin{tabular}{p{4.19em}cccccccc}
    \hline
    \specialrule{0em}{0.5pt}{0.5pt}
    \hline
    \multicolumn{1}{c}{} & Channel1 & Channel2 & Channel3 & Channel4 & Channel5 & Channel6 & Channel7 & Channel8 \\
    \midrule
    $SNR_B$ & 2.81  & 3.87  & 7.25  & 5.99  & 7.37  & 5.92  & 6.13  & 6.35 \\
    $SNR_H$ & 0.99  & 0.16  & 1.01  & 1.08  & 5.38  & 2.09  & 1.29  & 2.11 \\
    \hline
    \specialrule{0em}{0.5pt}{0.5pt}
    \hline
    \end{tabular}%
  \label{tab:addlabel}%
\end{table*}%

\begin{table*}[htbp]
  \centering
  \caption{$SNR_B$ and  $SNR_H$ average values in all areas.}
    \begin{tabular}{p{4.19em}cccccccc}
    \hline
    \specialrule{0em}{0.5pt}{0.5pt}
    \hline
    \multicolumn{1}{c}{} & Channel1 & Channel2 & Channel3 & Channel4 & Channel5 & Channel6 & Channel7 & Channel8 \\
    \midrule
    $SNR_B$ & 8.91  & 8.31  & 8.86  & 8.7   & 8.58  & 8.8   & 8.97  & 7.76 \\
    $SNR_H$ & 2.54  & 1.84  & 2.29  & -0.32 & 2.11  & 3.07  & 3.48  & 3.96 \\
    \hline
    \specialrule{0em}{0.5pt}{0.5pt}
    \hline
    \end{tabular}%
  \label{tab:addlabel}%
\end{table*}%

\begin{table*}[htbp]
  \centering
  \caption{$SNR_B$ and  $SNR_H$ average values for left half area.}
    \begin{tabular}{p{4.19em}cccccccc}
     \hline
    \specialrule{0em}{0.5pt}{0.5pt}
    \hline
    \multicolumn{1}{c}{} & Channel1 & Channel2 & Channel3 & Channel4 & Channel5 & Channel6 & Channel7 & Channel8 \\
    \midrule
    $SNR_B$ & 4.42  & 5.78  & 9.79  & 7.17  & 7.99  & 6.37  & 0.82  & 3.03 \\
    $SNR_H$ & 1.54  & 2.21  & 2.74  & 4.77  & 3.06  & 4.93  & 2.6   & -1.02 \\
    \hline
    \specialrule{0em}{0.5pt}{0.5pt}
    \hline
    \end{tabular}%
  \label{tab:addlabel}%
\end{table*}%

\section{RESULTS AND DISCUSSION}
This section systematically analyzes the experimental results from three dimensions. Firstly, the existence of organ radiation spatial distribution characteristics is verified and its impact is analyzed (Part A). Secondly, Taking one set of data as an example, we demonstrate the signal processing process of the method proposed in this paper. The detection performance of the three schemes under different scenarios is compared (Part B). Finally, the physical mechanisms responsible for the performance improvement are analyzed in depth (Part C). All experiments are based on the strap sensor data as the ground truth.

\subsection{Validation and analysis of spatial distribution characteristics of organ radiation }
To verify the existence of spatial distribution characteristics of organ radiation, experiments sequentially irradiated three effective areas: the right half area, the entire areas, and the left half area. 

The average $SNR_B$ and $SNR_H$ of the four subjects were calculated for each of the eight radar channels. The results are shown in Table $\mathrm{II}$, Table $\mathrm{III}$ and Table $\mathrm{IV}$. Based on the table data, the corresponding rules can be obtained.

\begin{figure}[htbp]
    \centering
    \includegraphics[width=0.5\textwidth]{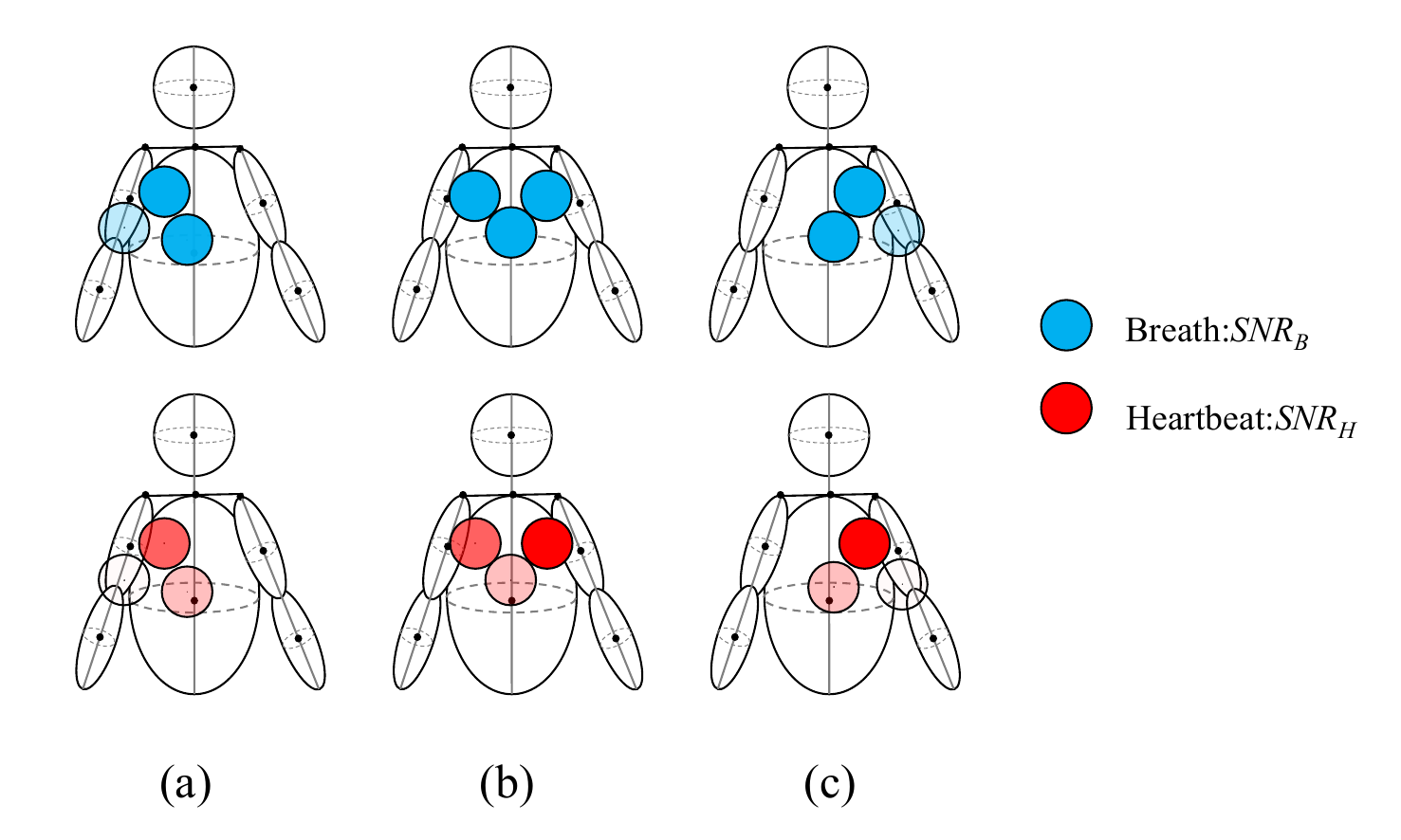} 
    \caption{$SNR_B$ and $SNR_H$ distribution (front view). (a) Left half area. (b) All areas. (c) Right half area.}

  \label{fig7}
  \end{figure}
\subsubsection{Right Half Area Irradiation} (Table $\mathrm{II}$). \textit{a)} Channels 1-2 exhibited low $SNR_B$ (2.81, 3.87) and $SNR_H$ (0.99, 0.16), corresponding to the right arm, where cardiopulmonary signals are weak. \textit{b)} Channels 3-6 showed high $SNR_B$ (7.25, 5.99, 7.37, 5.92) and $SNR_H$ (1.01, 1.08, 5.38, 2.09), indicating optimal detection at the 4th to 5th intercostal space on the right chest. \textit{c)} Channels 7-8 had high $SNR_B$ (6.13, 6.35) but low $SNR_H$ (1.29, 2.11), suggesting that respiration is detectable, but the heartbeat signal is weak at the 6th to 7th intercostal space on the right chest.

\subsubsection{All Areas Irradiation} (Table $\mathrm{III}$). \textit{a)}	All channels achieved high $SNR_B$, reflecting robust detection of respiration across the entire thoracoabdominal region. \textit{b)} $SNR_H$ was low in the central channels (e.g., channel 4: -0.32), likely due to overlap with the primary respiratory energy band (from the sternal angle to the xiphoid), where cardiac signals are attenuated. \textit{c)} Peripheral channels (e.g., channels 6-8) showed higher $SNR_H$ (3.07, 3.48, 3.96), indicating stronger heartbeat signals outside the respiratory-dominant zone.

\begin{figure}[htbp]
    \centering
    \includegraphics[width=0.5\textwidth]{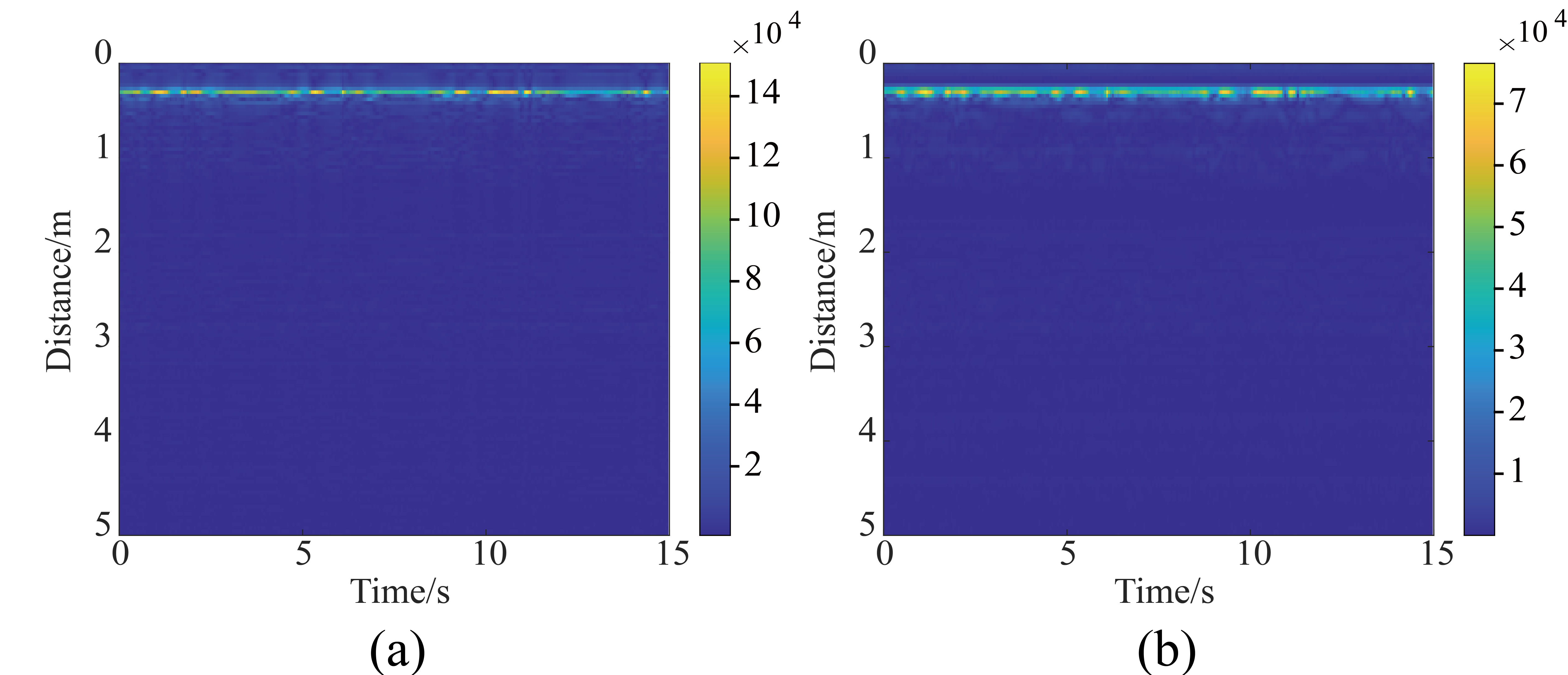} 
    \caption{Echo signal preprocessing. (a) Time-range map of original signal. (b) The preprocessed time-range map.}

  \label{fig8}
  \end{figure}

\begin{figure}[htbp]
    \centering
    \includegraphics[width=0.5\textwidth]{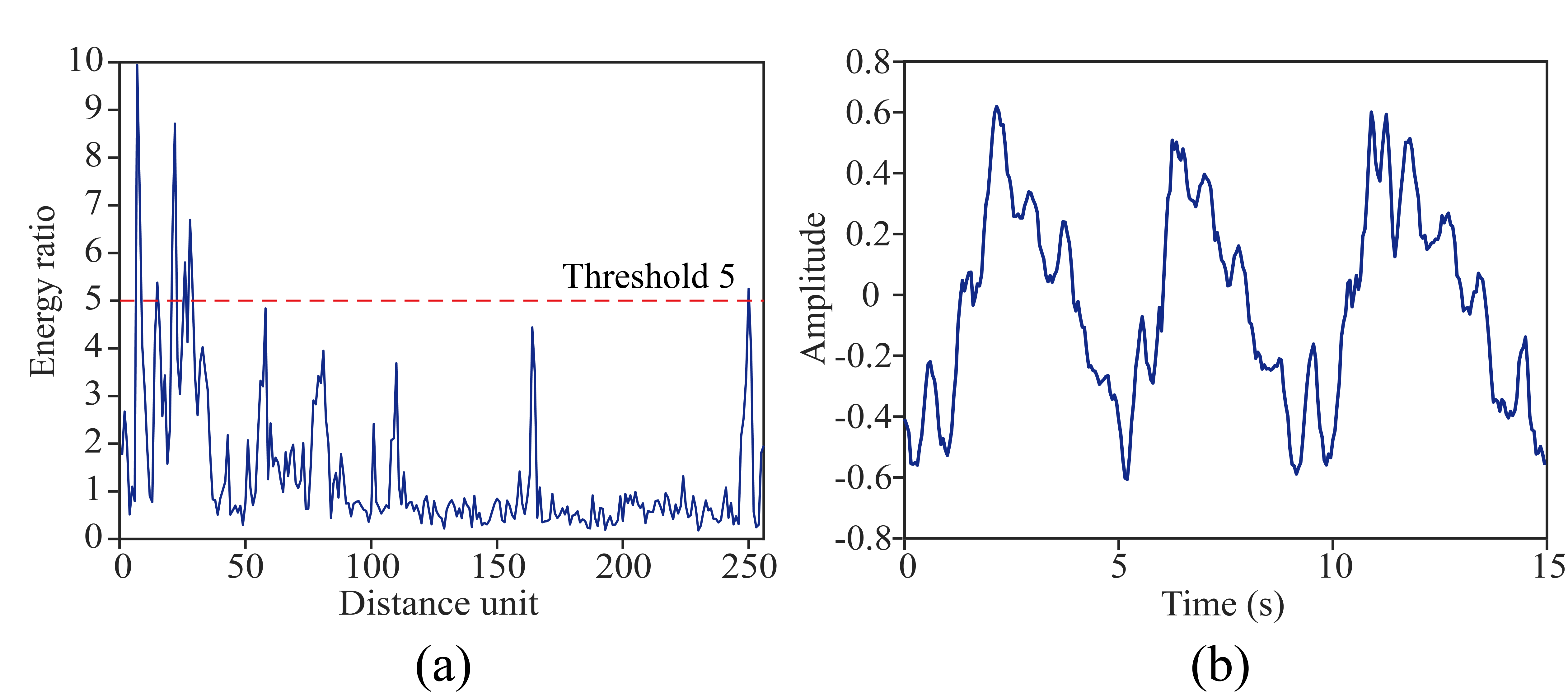} 
    \caption{(a) Energy ratio of different range bins. (b) The result of coherent fusion of multiple effective range bins signals.}

  \label{fig9}
  \end{figure}

\subsubsection{Left Half Area Irradiation} (Table $\mathrm{IV}$). \textit{a)} The results of irradiation in the left half mirrored the right half symmetrically. \textit{b)} Channels 7-8 (left arm analog) had low $SNR_B$ (0.82, 3.03) and $SNR_H$ (2.60, -1.02). \textit{c)} Channels 3-6 (left chest 4-5th intercostal) showed high $SNR_B$ (9.79, 7.17, 7.99, 6.37) and $SNR_H$ (2.74, 4.77, 3.06, 4.93). \textit{d)} Channels 1-2 (left chest 6-7th intercostal) had high $SNR_B$ (4.42, 5.78) but low $SNR_H$ (1.54, 2.21).

Based on the experimental results, we plotted the distribution of $SNR_B$ and $SNR_H$ across the thoracoabdominal region. As shown in Fig. 6 (all are front views), the first row represents the distribution of respiratory $SNR_B$, and the second row represents the distribution of heartbeat $SNR_H$. Its color depth represents the size of SNR, the darker the color, the greater the SNR. As seen in the figure, the main respiratory energy zone (blue) is located in the longitudinal band from the sternal angle to the xiphoid process, while the main heartbeat energy zone (red) is in the apical region between the left chest 4th and 5th ribs. This spatial distribution aligns with the analysis of the organ radiation spatial distribution characteristics in Section III, Part C.


\begin{figure}[htbp]
    \centering
    \includegraphics[width=0.5\textwidth]{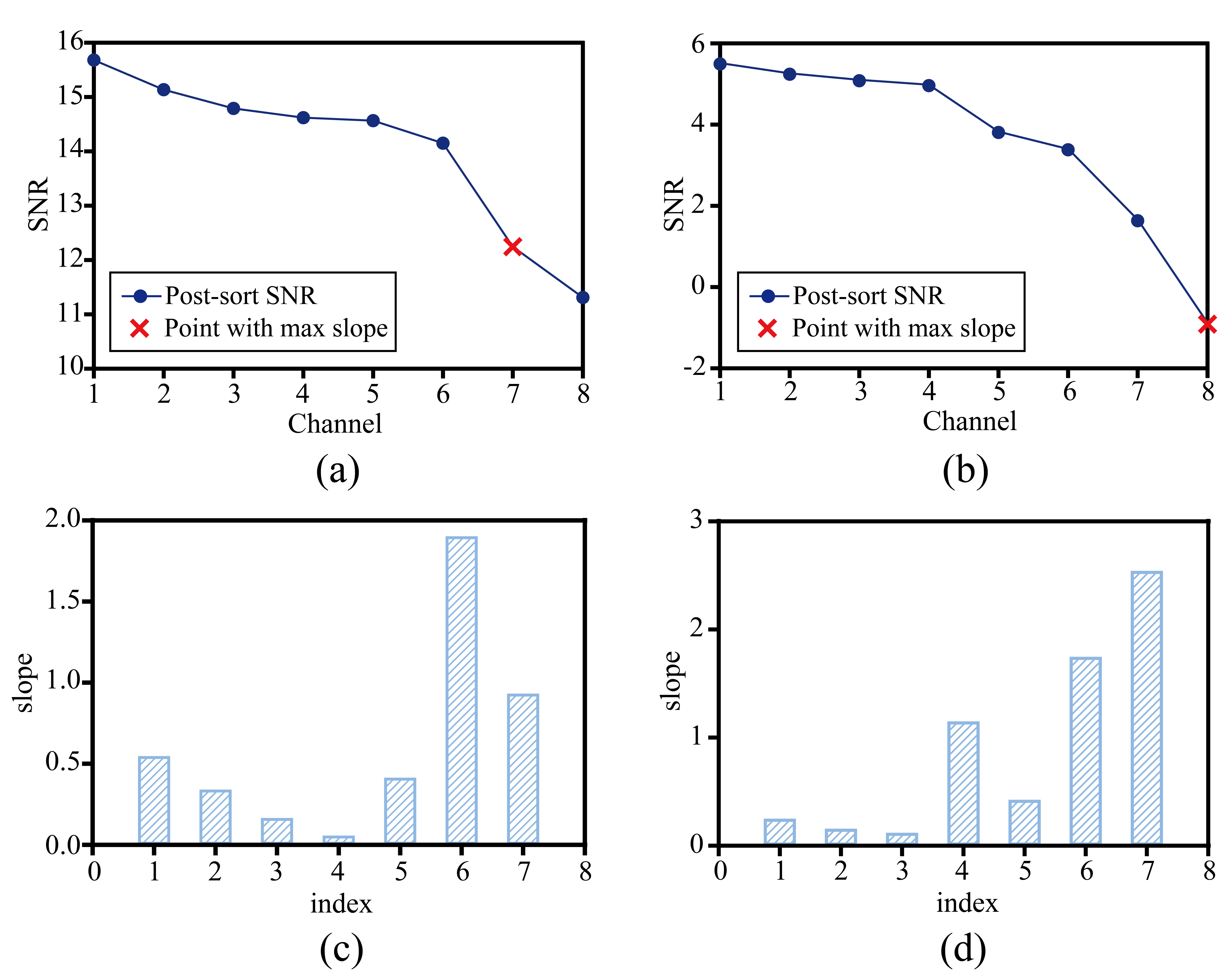} 
    \caption{Sort $SNR_B$ and $SNR_H$ in descending order and identify significant slope change points using first-order difference. (a) The descending order results of $SNR_B$ for each channel. (b) The descending order results of $SNR_H$ for each channel. (c) The first-order difference results of $SNR_B$. (d) The first-order difference results of $SNR_H$.}

  \label{fig10}
  \end{figure}

\begin{figure}[htbp]
    \centering
    \includegraphics[width=0.5\textwidth]{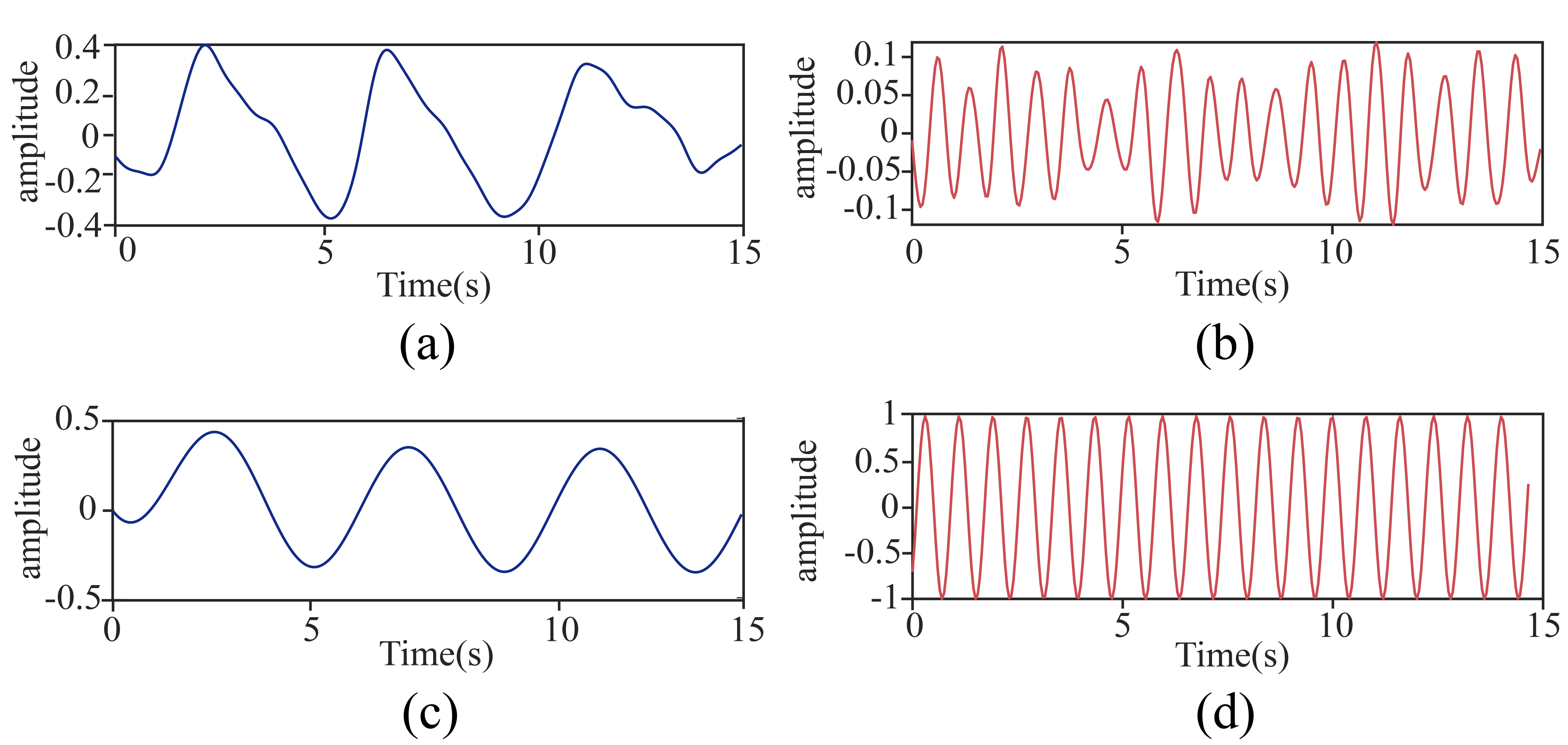} 
    \caption{Optimal template matching. (a) The weighted fusion results of the respiratory signals from the respiratory-dominant and mixed channels. (b) The weighted fusion results of the heartbeat signals from the heartbeat-dominant and mixed channels. (c) The optimal template matching results of the respiratory signals. (d) The optimal template matching results of the heartbeat signals.}

  \label{fig11}
  \end{figure}
\subsection{Algorithm performance verification}
To systematically evaluate the detection accuracy and robustness of the method proposed in this paper under different detection distances and human states, a comprehensive analysis of the detection error rates of RR and HR was conducted based on experimental data from 8 subjects. The experiments were carried out under three distance conditions (0.3 m, 1 m, and 2 m) and four human states (SI, ST, LS, and HT). To demonstrate the superiority of the proposed method, it was compared with Ref1 and Ref2 across all scenarios. The results measured by the respiratory and heartbeat strap, the RR and HR values estimated by the three methods, and the error rates of the three methods for estimating RR and HR are all presented in the appendix in the form of tables. The results are presented in the main text in the form of box plots and heatmaps.

\subsubsection{A case study}
\
\newline
\indent According to the signal processing flow of our proposed method, first, the echo signals are preprocessed and effective range bins are selected, with Channel 1 used as an example to demonstrate the algorithm's performance. Fig. 7 (a) shows the time-range map of the original echo signal. After preprocessing the original echo signal, the result is shown in Fig. 7 (b). It can be observed that after DC removal and Blackman window filtering, the background noise is effectively suppressed, and the chest wall micro-motion related signals present a clearer time-varying trajectory. The energy ratio of different range bins is shown in Fig. 8 (a). We select the range bins where $E_{in}/E_{out}>5$ as effective range bins and perform coherent fusion of signals from multiple effective range bins, with the result shown in Fig. 8 (b). Then, based on the spatial distribution characteristics of organ radiation, the multi-channel echoes obtained from MIMO radar detection are processed. The SNR for respiration/heartbeat in each channel is calculated and sorted in descending order, with the results shown in Fig. 9 (a) and Fig. 9 (b). Fig. 9 (c) and Fig. 9 (d) display the first-order differences of the respiration/heartbeat SNR. The respiration/heartbeat signals after principal component division and multi-channel signal weighted fusion are shown in Fig. 10 (a) and Fig. 10 (b), with their best-matched template signals shown in Fig. 10 (c) and Fig. 10 (d).

\begin{figure*}[htbp]
    \centering
    \includegraphics[width=\textwidth]{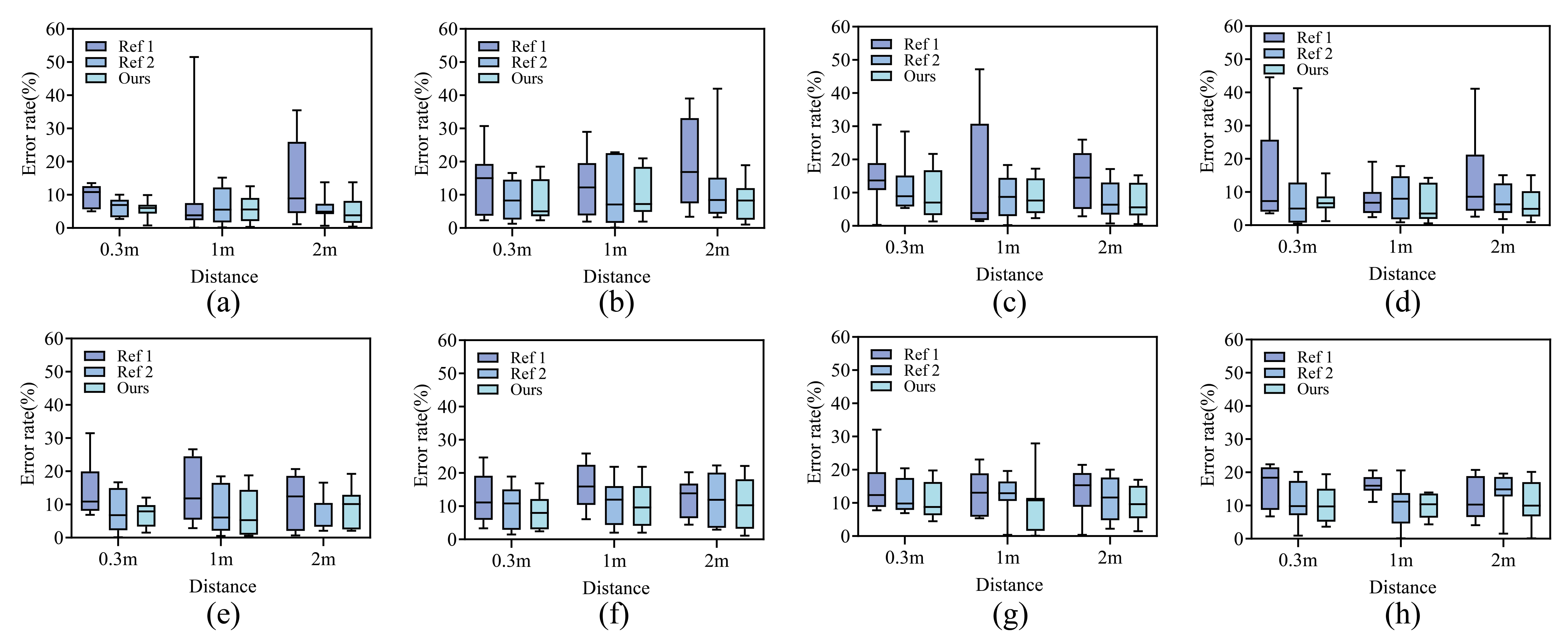} 
    \caption{The impact of distance on physiological parameter extraction performance. The first row is the error rate of estimating RR, and the second row is the error rate of estimating HR. Each subgraph represents a different state. (a) SI RR. (b) ST RR. (c) LS RR. (d) HT RR. (e) SI HR. (f) ST HR. (g) LS HR. (h) HT HR.}

  \label{fig11}
  \end{figure*}
  \begin{figure*}[htbp]
    \centering
    \includegraphics[width=\textwidth]{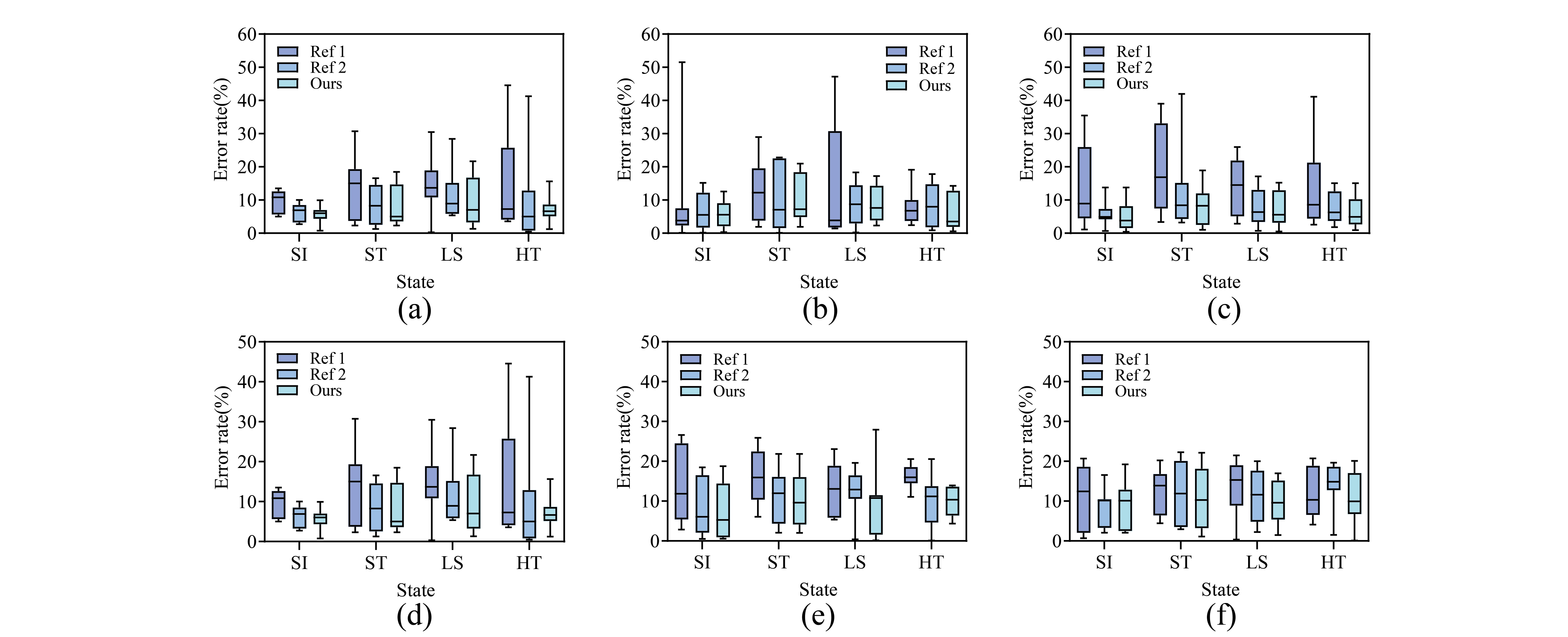} 
    \caption{The impact of state on physiological parameter extraction performance. The first row is the error rate of estimating RR, and the second row is the error rate of estimating HR. (a) 0.3 m RR. (b) 1 m RR. (c) 2 m RR. (d) 0.3 m HR. (e) 1 m HR. (f) 2 m HR.}

  \label{fig12}
  \end{figure*}

\subsubsection{The impact of distance on physiological parameter extraction performance}
\
\newline
\indent To investigate the impact of distance on the proposed method, the data was first visualized from the perspective of distance. Fig. 11 presents the test results of all subjects under different distances. Among the three methods, Ref 1 exhibits a high degree of data dispersion (characterized by a longer box), making it prone to errors due to distance changes or individual differences. This phenomenon is particularly evident in the estimation of RR. Taking Fig. 11 (a) as an example, the maximum error rate of Ref 1 at 0.3 m is only 13.54\%, while it reaches as high as 51.53\% at 1 m. In addition, at 1 m, the difference between the maximum and minimum error rates of Ref 1 is 51.38\%. Ref 2 shows moderate performance, but occasionally experiences an increase in the fluctuation range of errors due to distance changes. Our proposed method demonstrates a low and stable median as well as a smaller interquartile range in all subgraphs, indicating insensitivity to distance variations. It has a narrow fluctuation range and exhibits good adaptability to individual differences.

\begin{figure*}[htbp]
    \centering
    \includegraphics[width=\textwidth]{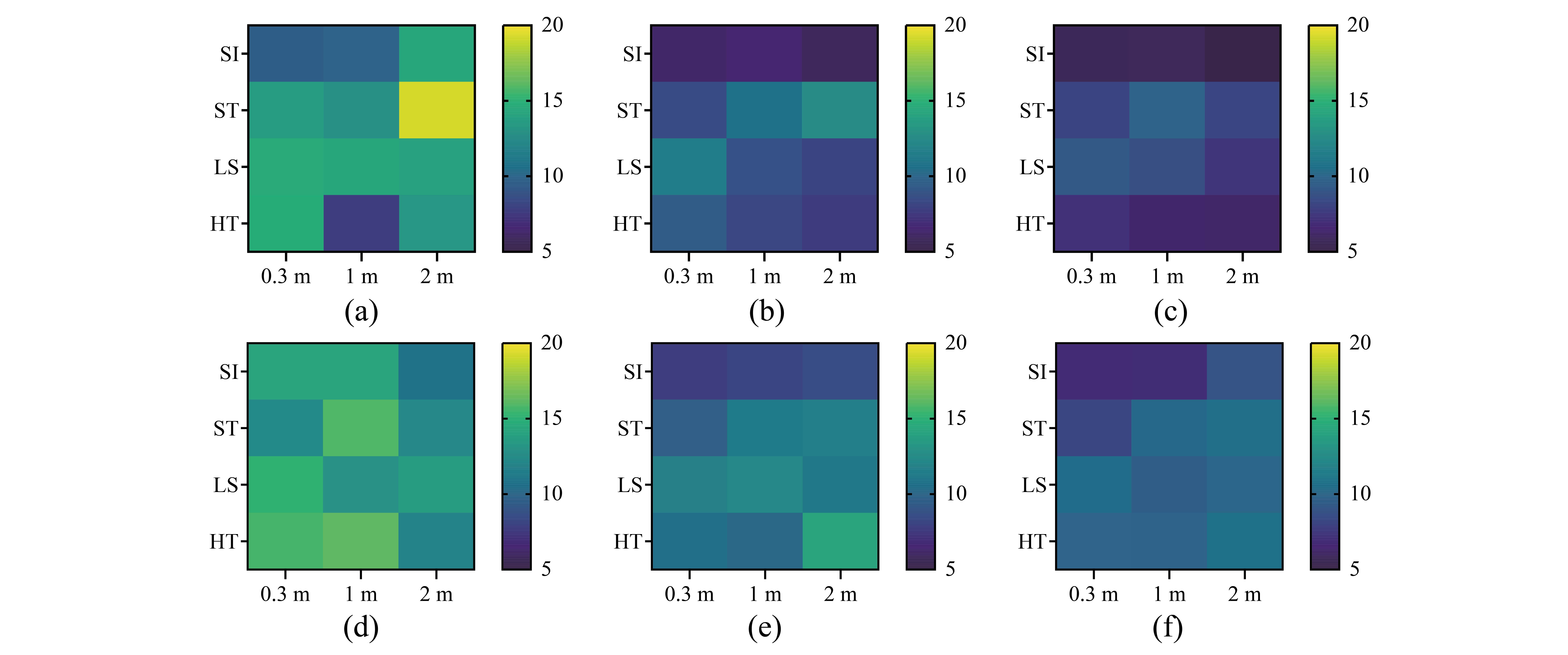} 
    \caption{The impact of state on physiological parameter extraction performance The first row is the error rate of estimating RR, and the second row is the error rate of estimating HR. (a) Ref1 RR. (b) Ref2 RR. (c) Ours RR. (d) Ref1 HR. (e) Ref2 HR. (f) Ours HR.}

  \label{fig12}
  \end{figure*}
\subsubsection{The impact of state on physiological parameter extraction performance}
\
\newline
\indent To investigate the impact of state variations, we visualized the data from the perspective of states. As shown in Fig. 12, box plots are used to present the test results of all subjects under different states. From the statistical results of the box plots, it can be clearly observed that Ref 1 is highly prone to significant fluctuations in data dispersion when the state changes. For example, in Fig. 12 (a), under the SI state, the difference between the maximum and minimum error rates of Ref 1 in RR estimation is 8.51\%, but when the state changes to ST, this difference suddenly increases to 28.41\%. Such fluctuations occur moderately in Ref 2. In most of the scenarios explored in this paper, the fluctuation range of error rates in RR and HR estimation for different individuals using our proposed method is relatively smaller compared to Ref 1 and Ref 2, indicating that our proposed method is insensitive to state changes.

\subsubsection{Analysis of Heatmaps for Distance-State Coupling Effects}
\
\newline
\indent The heatmaps clearly reveal the performance distribution pattern under the distance-state coupling effect. Overall, the proposed method achieves the best performance across all 12 scenarios, while Ref 1 performs the worst. 

In micro-motion scenarios, Ref 1 exhibits the poorest performance. Specifically, in short-distance detection scenarios ($\le $ 1 m) when estimating HR, only the results under (0.3 m, ST) and (1 m, LS) are relatively favorable. This outcome is consistent with the analysis in the introduction regarding the limitations of current single-channel radar physiological signal detection methods, which are particularly ineffective in short-distance detection and lack adaptability to dynamic scenarios.

When detection conditions change, our proposed method demonstrates the optimal robustness, as reflected by the uniform color scale distribution in its heatmap without drastic color variations. Due to its insufficient utilization of spatial information, Ref 2 exhibits sudden color block changes in the heatmap (e.g., the color suddenly shifts to light green in RR estimation under the (2 m, ST) condition and HR estimation under the (2 m, HT) condition), indicating slightly inferior robustness. Ref1 exhibits the poorest robustness, when estimating RR, it shows a sudden shift to yellow in the (2 m, SI) scenario and a drastic change to dark blue in the (1 m, HT) scenario.

The essential reason for this performance discrepancy lies in the following: Our proposed method, through multi-channel fusion guided by the spatial distribution characteristics of organ radiation, can stably capture signals from the main energy regions of respiration/heartbeat even in long-distance micro-motion scenarios. However, due to spatial limitations, the reference methods fail to continuously lock onto the optimal detection region under the same scenarios, leading to degraded robustness.

\subsection{Performance analysis}
The proposed method shows a lower error rate compared to the reference method. The fundamental reason is that the proposed approach leverages the spatial distribution characteristics of organ radiation. It classifies the dominant components and removes irrelevant signals, then performs weighted fusion based on the corresponding SNR. This process helps suppress motion interference in ineffective regions and reduces mutual interference between cardiopulmonary organs, thereby improving the quality of respiration and heartbeat detection. Moreover, the use of template matching to extract RR and HR also helps reduce high-frequency noise interference, further enhancing the accuracy of RR and HR estimation.

\section{Conclusion}
This paper proposes a physiological sign enhancement detection method for MIMO FMCW area array radar. Through multi-stage physiological driving fusion detection such as multi-scattering point modeling, multi-range unit coherent fusion in single channel and multi-channel fusion based on spatial distribution characteristics of organ radiation, the accuracy and robustness of respiratory heartbeat detection are significantly improved. The results show that our method outperforms the reference methods and exhibits strong robustness when scenarios change. This study promote the application of non-contact vital sign detection technology in complex and dynamic real-world scenarios. In the future, we will continue to conduct in-depth research on non-contact vital sign detection technology, promoting its evolution from rough sign detection to refined respiratory and heartbeat rhythm detection.
\appendices
\section*{Appendix A}
\renewcommand{\thetable}{A\arabic{table}}
\setcounter{table}{0}

\begin{table*}[htbp]
  \centering
  \caption{Experimental Results under the (0.3 m, SI) Scenario}
    \begin{tabular}{ccccccccccccccc}
    \hline
    \specialrule{0em}{0.5pt}{0.5pt}
    \hline
    \multirow{2}[4]{*}{subject ID} & \multicolumn{2}{c}{True value (bpm)} & \multicolumn{3}{c}{Estimated RR (bpm)} & \multicolumn{3}{c}{Estimated HR (bpm)} & \multicolumn{3}{c}{RR Error rate(\%)} & \multicolumn{3}{c}{HR Error rate (\%)} \\
\cmidrule{2-15}          & Chest Strap RR & Chest Strap HR & Ref1  & Ref2  & Ours  & Ref1  & Ref2  & Ours  & Ref1  & Ref2  & Ours  & Ref1  & Ref2  & Ours \\
    \midrule
    1     & 13.10  & 74.00  & 13.78  & 12.20  & 13.20  & 68.92  & 74.10  & 76.40  & 5.19  & 6.87  & 0.76  & 6.86  & 0.14  & 3.24  \\
    2     & 13.45  & 73.00  & 12.00  & 12.10  & 12.90  & 79.17  & 70.80  & 80.40  & 10.78  & 10.04  & 4.09  & 8.45  & 3.01  & 10.14  \\
    3     & 13.47  & 86.00  & 12.00  & 12.30  & 12.60  & 79.17  & 84.30  & 84.70  & 10.91  & 8.69  & 6.46  & 7.94  & 1.98  & 1.51  \\
    4     & 14.21  & 80.00  & 15.83  & 14.70  & 15.00  & 68.12  & 88.20  & 87.10  & 11.40  & 3.45  & 5.56  & 14.85  & 10.25  & 8.87  \\
    5     & 12.89  & 76.00  & 12.00  & 13.30  & 12.00  & 69.00  & 82.80  & 82.80  & 6.90  & 3.18  & 6.90  & 9.21  & 8.95  & 8.95  \\
    6     & 15.71  & 83.00  & 16.50  & 16.80  & 16.50  & 72.60  & 79.20  & 80.10  & 5.03  & 6.94  & 5.03  & 12.53  & 4.58  & 3.49  \\
    7     & 15.01  & 88.00  & 16.97  & 16.20  & 16.50  & 68.92  & 73.30  & 77.40  & 13.06  & 7.93  & 9.93  & 21.68  & 16.70  & 12.05  \\
    8     & 13.88  & 70.00  & 12.00  & 13.50  & 12.90  & 92.02  & 81.60  & 74.90  & 13.54  & 2.74  & 7.06  & 31.46  & 16.57  & 7.00  \\
    \hline
    \specialrule{0em}{0.5pt}{0.5pt}
    \hline
    \end{tabular}%
  \label{tab:addlabel}%
\end{table*}%

\begin{table*}[htbp]
  \centering
  \caption{Experimental Results under the (0.3 m, ST) Scenario}
    \begin{tabular}{ccccccccccccccc}
    \hline
    \specialrule{0em}{0.5pt}{0.5pt}
    \hline
    \multirow{2}[4]{*}{subject ID} & \multicolumn{2}{c}{True value (bpm)} & \multicolumn{3}{c}{Estimated RR (bpm)} & \multicolumn{3}{c}{Estimated HR (bpm)} & \multicolumn{3}{c}{RR Error rate (\%)} & \multicolumn{3}{c}{HR Error rate (\%)} \\
\cmidrule{2-15}          & Chest Strap RR & Chest Strap HR & Ref1  & Ref2  & Ours  & Ref1  & Ref2  & Ours  & Ref1  & Ref2  & Ours  & Ref1  & Ref2  & Ours \\
    \midrule
    1     & 14.33  & 83.00  & 12.40  & 15.30  & 15.00  & 72.00  & 98.70  & 81.00  & 13.47  & 6.77  & 4.68  & 13.25  & 18.92  & 2.41  \\
    2     & 13.21  & 82.00  & 10.80  & 11.20  & 12.00  & 72.30  & 83.20  & 76.50  & 18.24  & 15.22  & 9.16  & 11.83  & 1.46  & 6.71  \\
    3     & 13.64  & 92.00  & 15.90  & 15.90  & 15.90  & 69.30  & 80.20  & 89.10  & 16.57  & 16.57  & 16.57  & 24.67  & 12.83  & 3.15  \\
    4     & 17.32  & 87.00  & 12.00  & 17.10  & 18.10  & 78.60  & 73.50  & 72.30  & 30.72  & 1.27  & 4.50  & 9.66  & 15.52  & 16.90  \\
    5     & 16.12  & 90.00  & 15.30  & 17.70  & 19.10  & 87.00  & 87.00  & 87.30  & 5.09  & 9.80  & 18.49  & 3.33  & 3.33  & 3.00  \\
    6     & 17.30  & 89.00  & 17.70  & 17.70  & 17.70  & 70.20  & 76.80  & 78.30  & 2.31  & 2.31  & 2.31  & 21.12  & 13.71  & 12.02  \\
    7     & 14.53  & 85.00  & 15.00  & 15.00  & 15.00  & 88.90  & 82.70  & 95.40  & 3.23  & 3.23  & 3.23  & 4.59  & 2.71  & 12.24  \\
    8     & 13.95  & 78.00  & 16.70  & 15.70  & 14.70  & 69.90  & 84.90  & 85.20  & 19.71  & 12.54  & 5.38  & 10.38  & 8.85  & 9.23  \\
    \hline
    \specialrule{0em}{0.5pt}{0.5pt}
    \hline
    \end{tabular}%
  \label{tab:addlabel}%
\end{table*}%

\begin{table*}[htbp]
  \centering
  \caption{Experimental Results under the (0.3 m, LS) Scenario}
    \begin{tabular}{ccccccccccccccc}
    \hline
    \specialrule{0em}{0.5pt}{0.5pt}
    \hline
    \multirow{2}[4]{*}{subject ID} & \multicolumn{2}{c}{True value (bpm)} & \multicolumn{3}{c}{Estimated RR (bpm)} & \multicolumn{3}{c}{Estimated HR (bpm)} & \multicolumn{3}{c}{RR Error rate (\%)} & \multicolumn{3}{c}{HR Error rate (\%)} \\
\cmidrule{2-15}          & Chest Strap RR & Chest Strap HR & Ref1  & Ref2  & Ours  & Ref1  & Ref2  & Ours  & Ref1  & Ref2  & Ours  & Ref1  & Ref2  & Ours \\
    \midrule
    1     & 16.75  & 79.00  & 16.80  & 15.80  & 17.40  & 72.30  & 71.40  & 74.40  & 0.30  & 5.67  & 3.88  & 8.48  & 9.62  & 5.82  \\
    2     & 17.92  & 76.00  & 19.80  & 20.80  & 19.20  & 90.30  & 67.90  & 72.60  & 10.49  & 16.07  & 7.14  & 18.82  & 10.66  & 4.47  \\
    3     & 16.11  & 72.00  & 11.20  & 14.90  & 15.90  & 95.10  & 78.90  & 79.50  & 30.48  & 7.51  & 1.30  & 32.08  & 9.58  & 10.42  \\
    4     & 19.73  & 85.00  & 22.50  & 22.20  & 22.20  & 75.00  & 76.50  & 77.40  & 14.04  & 12.52  & 12.52  & 11.76  & 10.00  & 8.94  \\
    5     & 20.31  & 78.00  & 17.60  & 18.20  & 18.90  & 85.60  & 72.60  & 72.00  & 13.34  & 10.39  & 6.94  & 9.74  & 6.92  & 7.69  \\
    6     & 13.41  & 71.00  & 10.80  & 9.60  & 10.50  & 76.50  & 76.20  & 84.00  & 19.46  & 28.41  & 21.70  & 7.75  & 7.32  & 18.31  \\
    7     & 20.31  & 89.00  & 16.80  & 21.40  & 24.00  & 71.70  & 70.80  & 81.30  & 17.28  & 5.37  & 18.17  & 19.44  & 20.45  & 8.65  \\
    8     & 18.35  & 95.00  & 16.20  & 19.50  & 18.90  & 82.70  & 76.20  & 76.20  & 11.72  & 6.27  & 3.00  & 12.95  & 19.79  & 19.79  \\
    \hline
    \specialrule{0em}{0.5pt}{0.5pt}
    \hline
    \end{tabular}%
  \label{tab:addlabel}%
\end{table*}%

\begin{table*}[htbp]
  \centering
  \caption{Experimental Results under the (0.3 m, HT) Scenario}
    \begin{tabular}{ccccccccccccccc}
    \hline
    \specialrule{0em}{0.5pt}{0.5pt}
    \hline
    \multirow{2}[4]{*}{subject ID} & \multicolumn{2}{c}{True value (bpm)} & \multicolumn{3}{c}{Estimated RR (bpm)} & \multicolumn{3}{c}{Estimated HR (bpm)} & \multicolumn{3}{c}{RR Error rate (\%)} & \multicolumn{3}{c}{HR Error rate (\%)} \\
\cmidrule{2-15}          & Chest Strap RR & Chest Strap HR & Ref1  & Ref2  & Ours  & Ref1  & Ref2  & Ours  & Ref1  & Ref2  & Ours  & Ref1  & Ref2  & Ours \\
    \midrule
    1     & 19.93  & 86.00  & 20.70  & 11.70  & 20.90  & 105.30  & 92.70  & 89.10  & 3.86  & 41.29  & 4.87  & 22.44  & 7.79  & 3.60  \\
    2     & 13.75  & 85.00  & 9.90  & 11.70  & 15.90  & 92.70  & 79.20  & 80.40  & 28.00  & 14.91  & 15.64  & 9.06  & 6.82  & 5.41  \\
    3     & 13.69  & 84.00  & 13.20  & 13.20  & 12.90  & 76.80  & 75.30  & 75.60  & 3.58  & 3.58  & 5.77  & 8.57  & 10.36  & 10.00  \\
    4     & 18.92  & 89.00  & 15.30  & 17.70  & 17.70  & 72.00  & 71.10  & 71.70  & 19.13  & 6.45  & 6.45  & 19.10  & 20.11  & 19.44  \\
    5     & 16.58  & 98.00  & 15.30  & 16.50  & 18.10  & 80.70  & 88.20  & 88.00  & 7.72  & 0.48  & 9.17  & 17.65  & 10.00  & 10.20  \\
    6     & 20.35  & 86.00  & 21.30  & 20.70  & 20.10  & 80.20  & 85.20  & 81.70  & 4.67  & 1.72  & 1.23  & 6.74  & 0.93  & 5.00  \\
    7     & 23.81  & 98.00  & 13.20  & 23.70  & 22.10  & 78.60  & 78.60  & 81.60  & 44.56  & 0.46  & 7.18  & 19.80  & 19.80  & 16.73  \\
    8     & 9.66  & 89.00  & 9.00  & 9.00  & 9.00  & 69.30  & 80.40  & 80.60  & 6.83  & 6.83  & 6.83  & 22.13  & 9.66  & 9.44  \\
    \hline
    \specialrule{0em}{0.5pt}{0.5pt}
    \hline
    \end{tabular}%
  \label{tab:addlabel}%
\end{table*}%


\begin{table*}[htbp]
  \centering
  \caption{Experimental Results under the (1 m, SI) Scenario}
    \begin{tabular}{ccccccccccccccc}
    \hline
    \specialrule{0em}{0.5pt}{0.5pt}
    \hline
    \multirow{2}[4]{*}{subject ID} & \multicolumn{2}{c}{True value (bpm)} & \multicolumn{3}{c}{Estimated RR (bpm)} & \multicolumn{3}{c}{Estimated HR (bpm)} & \multicolumn{3}{c}{RR Error rate (\%)} & \multicolumn{3}{c}{HR Error rate (\%)} \\
\cmidrule{2-15}          & Chest Strap RR & Chest Strap HR & Ref1  & Ref2  & Ours  & Ref1  & Ref2  & Ours  & Ref1  & Ref2  & Ours  & Ref1  & Ref2  & Ours \\
    \midrule
    1     & 17.51  & 85.00  & 18.90  & 19.90  & 18.90  & 72.80  & 83.40  & 84.30  & 7.94  & 13.65  & 7.94  & 14.35  & 1.88  & 0.82  \\
    2     & 13.79  & 88.00  & 13.50  & 14.70  & 14.40  & 69.90  & 78.20  & 88.50  & 2.10  & 6.60  & 4.42  & 20.57  & 11.14  & 0.57  \\
    3     & 19.21  & 87.00  & 19.80  & 16.30  & 16.80  & 84.50  & 83.70  & 83.10  & 3.07  & 15.15  & 12.55  & 2.87  & 3.79  & 4.48  \\
    4     & 21.37  & 94.00  & 20.70  & 21.00  & 21.30  & 69.00  & 94.50  & 87.60  & 3.14  & 1.73  & 0.33  & 26.60  & 0.53  & 6.81  \\
    5     & 20.07  & 89.00  & 20.10  & 20.40  & 20.40  & 80.70  & 105.30  & 72.30  & 0.15  & 1.64  & 1.64  & 9.33  & 18.31  & 18.76  \\
    6     & 18.57  & 83.00  & 9.00  & 18.60  & 19.20  & 86.40  & 85.20  & 78.00  & 51.53  & 0.16  & 3.39  & 4.10  & 2.65  & 6.02  \\
    7     & 19.46  & 106.00  & 20.70  & 21.00  & 21.30  & 78.60  & 86.40  & 88.00  & 6.37  & 7.91  & 9.46  & 25.85  & 18.49  & 16.98  \\
    8     & 13.19  & 91.00  & 12.60  & 12.60  & 12.30  & 82.50  & 83.40  & 90.00  & 4.47  & 4.47  & 6.75  & 9.34  & 8.35  & 1.10  \\
    \hline
    \specialrule{0em}{0.5pt}{0.5pt}
    \hline
    \end{tabular}%
  \label{tab:addlabel}%
\end{table*}%

\begin{table*}[htbp]
  \centering
  \caption{Experimental Results under the (1 m, ST) Scenario}
    \begin{tabular}{ccccccccccccccc}
    \hline
    \specialrule{0em}{0.5pt}{0.5pt}
    \hline
    \multirow{2}[4]{*}{subject ID} & \multicolumn{2}{c}{True value (bpm)} & \multicolumn{3}{c}{Estimated RR (bpm)} & \multicolumn{3}{c}{Estimated HR (bpm)} & \multicolumn{3}{c}{RR Error rate (\%)} & \multicolumn{3}{c}{HR Error rate (\%)} \\
\cmidrule{2-15}          & Chest Strap RR & Chest Strap HR & Ref1  & Ref2  & Ours  & Ref1  & Ref2  & Ours  & Ref1  & Ref2  & Ours  & Ref1  & Ref2  & Ours \\
    \midrule
    1     & 17.36  & 83.00  & 22.39  & 21.30  & 21.00  & 68.92  & 81.30  & 69.00  & 28.97  & 22.70  & 20.97  & 16.96  & 2.05  & 16.87  \\
    2     & 16.53  & 79.00  & 19.49  & 18.30  & 18.30  & 69.92  & 76.80  & 77.40  & 17.91  & 10.71  & 10.71  & 11.49  & 2.78  & 2.03  \\
    3     & 14.91  & 81.00  & 15.83  & 15.30  & 15.60  & 68.92  & 98.70  & 98.70  & 6.17  & 2.62  & 4.63  & 14.91  & 21.85  & 21.85  \\
    4     & 14.42  & 89.00  & 14.70  & 14.40  & 14.70  & 73.81  & 102.00  & 76.90  & 1.94  & 0.14  & 1.94  & 17.07  & 14.61  & 13.60  \\
    5     & 14.83  & 80.00  & 13.78  & 15.00  & 13.50  & 84.85  & 72.90  & 83.10  & 7.08  & 1.15  & 8.97  & 6.06  & 8.87  & 3.87  \\
    6     & 16.12  & 91.00  & 18.92  & 19.80  & 19.50  & 68.92  & 75.90  & 86.70  & 17.37  & 22.83  & 20.97  & 24.26  & 16.59  & 4.73  \\
    7     & 15.37  & 93.00  & 15.83  & 15.90  & 16.20  & 68.92  & 83.40  & 83.40  & 2.99  & 3.45  & 5.40  & 25.89  & 10.32  & 10.32  \\
    8     & 15.02  & 82.00  & 12.00  & 11.70  & 14.20  & 73.81  & 70.80  & 89.30  & 20.11  & 22.10  & 5.46  & 9.99  & 13.66  & 8.90  \\
    \hline
    \specialrule{0em}{0.5pt}{0.5pt}
    \hline
    \end{tabular}%
  \label{tab:addlabel}%
\end{table*}%

\begin{table*}[htbp]
  \centering
  \caption{Experimental Results under the (1 m, LS) Scenario}
    \begin{tabular}{ccccccccccccccc}
    \hline
    \specialrule{0em}{0.5pt}{0.5pt}
    \hline
    \multirow{2}[4]{*}{subject ID} & \multicolumn{2}{c}{True value (bpm)} & \multicolumn{3}{c}{Estimated RR (bpm)} & \multicolumn{3}{c}{Estimated HR (bpm)} & \multicolumn{3}{c}{RR Error rate (\%)} & \multicolumn{3}{c}{HR Error rate (\%)} \\
\cmidrule{2-15}          & Chest Strap RR & Chest Strap HR & Ref1  & Ref2  & Ours  & Ref1  & Ref2  & Ours  & Ref1  & Ref2  & Ours  & Ref1  & Ref2  & Ours \\
    \midrule
    1     & 14.88  & 79.00  & 9.80  & 13.00  & 13.10  & 94.50  & 94.50  & 80.10  & 34.14  & 12.63  & 11.96  & 19.62  & 19.62  & 1.39  \\
    2     & 15.87  & 75.00  & 16.50  & 16.90  & 15.50  & 84.30  & 84.30  & 83.40  & 3.97  & 6.49  & 2.33  & 12.40  & 12.40  & 11.20  \\
    3     & 16.13  & 82.00  & 15.90  & 15.30  & 15.20  & 86.40  & 93.00  & 91.40  & 1.43  & 5.15  & 5.77  & 5.37  & 13.41  & 11.46  \\
    4     & 17.14  & 96.00  & 16.50  & 14.00  & 16.50  & 82.80  & 85.20  & 84.90  & 3.73  & 18.32  & 3.73  & 13.75  & 11.25  & 11.56  \\
    5     & 17.85  & 97.00  & 17.40  & 19.80  & 17.10  & 80.70  & 80.10  & 69.90  & 2.52  & 10.92  & 4.20  & 16.80  & 17.42  & 27.94  \\
    6     & 17.79  & 95.00  & 9.40  & 17.40  & 16.10  & 100.20  & 85.20  & 85.20  & 47.16  & 2.19  & 9.50  & 5.47  & 10.32  & 10.32  \\
    7     & 19.55  & 98.00  & 15.50  & 22.50  & 22.50  & 75.40  & 84.60  & 96.00  & 20.72  & 15.09  & 15.09  & 23.06  & 13.67  & 2.04  \\
    8     & 18.34  & 78.00  & 18.60  & 18.30  & 21.50  & 72.60  & 78.30  & 78.10  & 1.42  & 0.22  & 17.23  & 6.92  & 0.38  & 0.13  \\
    \hline
    \specialrule{0em}{0.5pt}{0.5pt}
    \hline
    \end{tabular}%
  \label{tab:addlabel}%
\end{table*}%

\begin{table*}[htbp]
  \centering
  \caption{Experimental Results under the (1 m, HT) Scenario}
    \begin{tabular}{ccccccccccccccc}
    \hline
    \specialrule{0em}{0.5pt}{0.5pt}
    \hline
    \multirow{2}[4]{*}{subject ID} & \multicolumn{2}{c}{True value (bpm)} & \multicolumn{3}{c}{Estimated RR (bpm)} & \multicolumn{3}{c}{Estimated HR (bpm)} & \multicolumn{3}{c}{RR Error rate (\%)} & \multicolumn{3}{c}{HR Error rate (\%)} \\
\cmidrule{2-15}          & Chest Strap RR & Chest Strap HR & Ref1  & Ref2  & Ours  & Ref1  & Ref2  & Ours  & Ref1  & Ref2  & Ours  & Ref1  & Ref2  & Ours \\
    \midrule
    1     & 20.52  & 81.00  & 19.80  & 18.90  & 20.40  & 69.00  & 89.10  & 69.90  & 3.51  & 7.89  & 0.58  & 14.81  & 10.00  & 13.70  \\
    2     & 18.21  & 89.00  & 19.20  & 20.10  & 20.40  & 76.20  & 82.80  & 83.10  & 5.44  & 10.38  & 12.03  & 14.38  & 6.97  & 6.63  \\
    3     & 18.84  & 84.00  & 17.10  & 22.20  & 21.30  & 69.00  & 72.90  & 73.20  & 9.24  & 17.83  & 13.06  & 17.86  & 13.21  & 12.86  \\
    4     & 15.23  & 88.00  & 16.80  & 17.70  & 17.40  & 75.00  & 69.90  & 76.20  & 10.31  & 16.22  & 14.25  & 14.77  & 20.57  & 13.41  \\
    5     & 17.95  & 84.00  & 16.50  & 16.50  & 17.40  & 93.30  & 72.30  & 72.30  & 8.08  & 8.08  & 3.06  & 11.07  & 13.93  & 13.93  \\
    6     & 13.56  & 88.00  & 13.00  & 14.10  & 13.80  & 69.90  & 77.10  & 82.50  & 4.13  & 3.98  & 1.77  & 20.57  & 12.39  & 6.25  \\
    7     & 19.33  & 85.00  & 19.80  & 19.50  & 20.10  & 69.00  & 84.90  & 78.30  & 2.43  & 0.88  & 3.98  & 18.82  & 0.12  & 7.88  \\
    8     & 18.79  & 92.00  & 15.20  & 18.60  & 19.20  & 76.20  & 88.50  & 96.00  & 19.11  & 1.01  & 2.18  & 17.17  & 3.80  & 4.35  \\
    \hline
    \specialrule{0em}{0.5pt}{0.5pt}
    \hline
    \end{tabular}%
  \label{tab:addlabel}%
\end{table*}%

\begin{table*}[htbp]
  \centering
  \caption{Experimental Results under the (2 m, SI) Scenario}
    \begin{tabular}{ccccccccccccccc}
    \hline
    \specialrule{0em}{0.5pt}{0.5pt}
    \hline
    \multirow{2}[4]{*}{subject ID} & \multicolumn{2}{c}{True value (bpm)} & \multicolumn{3}{c}{Estimated RR (bpm)} & \multicolumn{3}{c}{Estimated HR (bpm)} & \multicolumn{3}{c}{RR Error rate (\%)} & \multicolumn{3}{c}{HR Error rate (\%)} \\
\cmidrule{2-15}          & Chest Strap RR & Chest Strap HR & Ref1  & Ref2  & Ours  & Ref1  & Ref2  & Ours  & Ref1  & Ref2  & Ours  & Ref1  & Ref2  & Ours \\
    \midrule
    1     & 16.27  & 84.00  & 10.50  & 15.60  & 16.20  & 86.40  & 75.30  & 82.20  & 35.46  & 4.12  & 0.43  & 2.86  & 10.36  & 2.14  \\
    2     & 17.98  & 87.00  & 17.30  & 15.50  & 15.50  & 69.00  & 88.80  & 84.00  & 3.78  & 13.79  & 13.79  & 20.69  & 2.07  & 3.45  \\
    3     & 20.46  & 86.00  & 19.10  & 19.50  & 19.50  & 74.10  & 91.50  & 74.40  & 6.65  & 4.69  & 4.69  & 13.84  & 6.40  & 13.49  \\
    4     & 19.29  & 91.00  & 14.10  & 18.30  & 19.80  & 89.40  & 81.90  & 73.50  & 26.91  & 5.13  & 2.64  & 1.76  & 10.00  & 19.23  \\
    5     & 12.86  & 90.00  & 11.70  & 12.00  & 11.70  & 72.60  & 80.70  & 81.60  & 9.02  & 6.69  & 9.02  & 19.33  & 10.33  & 9.33  \\
    6     & 15.14  & 87.00  & 13.80  & 14.00  & 14.70  & 86.40  & 72.60  & 85.20  & 8.85  & 7.53  & 2.91  & 0.69  & 16.55  & 2.07  \\
    7     & 16.69  & 79.00  & 16.50  & 16.80  & 16.50  & 92.10  & 87.30  & 87.60  & 1.14  & 0.66  & 1.14  & 16.58  & 10.51  & 10.89  \\
    8     & 13.79  & 85.00  & 10.60  & 13.20  & 13.00  & 75.60  & 83.10  & 75.50  & 23.13  & 4.28  & 5.73  & 11.06  & 2.24  & 11.18  \\
    \hline
    \specialrule{0em}{0.5pt}{0.5pt}
    \hline
    \end{tabular}%
  \label{tab:addlabel}%
\end{table*}%

\begin{table*}[htbp]
  \centering
  \caption{Experimental Results under the (2 m, ST) Scenario}
    \begin{tabular}{ccccccccccccccc}
    \hline
    \specialrule{0em}{0.5pt}{0.5pt}
    \hline
    \multirow{2}[4]{*}{subject ID} & \multicolumn{2}{c}{True value (bpm)} & \multicolumn{3}{c}{Estimated RR (bpm)} & \multicolumn{3}{c}{Estimated HR (bpm)} & \multicolumn{3}{c}{RR Error rate (\%)} & \multicolumn{3}{c}{HR Error rate (\%)} \\
\cmidrule{2-15}          & Chest Strap RR & Chest Strap HR & Ref1  & Ref2  & Ours  & Ref1  & Ref2  & Ours  & Ref1  & Ref2  & Ours  & Ref1  & Ref2  & Ours \\
    \midrule
    1     & 17.58  & 89.00  & 14.50  & 10.20  & 17.40  & 94.50  & 108.80  & 94.80  & 17.52  & 41.98  & 1.02  & 6.18  & 22.25  & 6.52  \\
    2     & 15.11  & 93.00  & 10.90  & 14.10  & 13.80  & 78.30  & 90.00  & 97.20  & 27.86  & 6.68  & 8.67  & 15.81  & 3.23  & 4.52  \\
    3     & 13.93  & 81.00  & 14.40  & 12.90  & 12.20  & 84.60  & 69.90  & 80.10  & 3.37  & 7.39  & 12.42  & 4.44  & 13.70  & 1.11  \\
    4     & 16.12  & 94.00  & 10.50  & 15.60  & 15.60  & 75.00  & 103.50  & 73.20  & 34.86  & 3.23  & 3.23  & 20.21  & 10.11  & 22.13  \\
    5     & 15.85  & 92.00  & 14.70  & 15.30  & 16.20  & 76.20  & 89.30  & 74.40  & 7.26  & 3.47  & 2.21  & 17.17  & 2.93  & 19.13  \\
    6     & 15.17  & 88.00  & 12.70  & 13.20  & 12.30  & 76.80  & 69.00  & 74.70  & 16.28  & 12.99  & 18.92  & 12.73  & 21.59  & 15.11  \\
    7     & 21.64  & 80.00  & 13.20  & 18.20  & 19.30  & 74.40  & 83.40  & 82.20  & 39.00  & 15.90  & 10.81  & 7.00  & 4.25  & 2.75  \\
    8     & 19.87  & 83.00  & 18.30  & 18.00  & 18.30  & 70.50  & 70.00  & 71.40  & 7.90  & 9.41  & 7.90  & 15.06  & 15.66  & 13.98  \\
    \hline
    \specialrule{0em}{0.5pt}{0.5pt}
    \hline
    \end{tabular}%
  \label{tab:addlabel}%
\end{table*}%

\begin{table*}[htbp]
  \centering
  \caption{Experimental Results under the (2 m, LS) Scenario}
    \begin{tabular}{ccccccccccccccc}
    \hline
    \specialrule{0em}{0.5pt}{0.5pt}
    \hline
    \multirow{2}[4]{*}{subject ID} & \multicolumn{2}{c}{True value (bpm)} & \multicolumn{3}{c}{Estimated RR (bpm)} & \multicolumn{3}{c}{Estimated HR (bpm)} & \multicolumn{3}{c}{RR Error rate (\%)} & \multicolumn{3}{c}{HR Error rate (\%)} \\
\cmidrule{2-15}          & Chest Strap RR & Chest Strap HR & Ref1  & Ref2  & Ours  & Ref1  & Ref2  & Ours  & Ref1  & Ref2  & Ours  & Ref1  & Ref2  & Ours \\
    \midrule
    1     & 15.88  & 88.00  & 15.00  & 14.70  & 14.10  & 75.30  & 82.80  & 84.30  & 5.54  & 7.43  & 11.21  & 14.43  & 5.91  & 4.20  \\
    2     & 16.72  & 89.00  & 15.90  & 16.20  & 15.90  & 72.00  & 81.30  & 90.30  & 4.90  & 3.11  & 4.90  & 19.10  & 8.65  & 1.46  \\
    3     & 17.01  & 89.00  & 14.30  & 16.10  & 17.10  & 69.90  & 72.60  & 73.90  & 15.93  & 5.35  & 0.53  & 21.46  & 18.43  & 16.97  \\
    4     & 22.56  & 84.00  & 16.70  & 21.60  & 19.50  & 74.10  & 71.10  & 71.10  & 25.98  & 4.26  & 13.56  & 11.79  & 15.36  & 15.36  \\
    5     & 15.41  & 81.00  & 12.10  & 15.30  & 15.00  & 81.30  & 79.20  & 73.80  & 21.48  & 0.71  & 2.66  & 0.37  & 2.22  & 8.89  \\
    6     & 17.92  & 79.00  & 17.40  & 15.60  & 16.80  & 85.20  & 82.50  & 72.00  & 2.90  & 12.95  & 6.25  & 7.85  & 4.43  & 8.86  \\
    7     & 14.32  & 85.00  & 16.20  & 16.20  & 16.50  & 69.00  & 72.60  & 72.60  & 13.13  & 13.13  & 15.22  & 18.82  & 14.59  & 14.59  \\
    8     & 18.10  & 87.00  & 14.10  & 15.00  & 17.30  & 72.90  & 69.60  & 78.00  & 22.10  & 17.13  & 4.42  & 16.21  & 20.00  & 10.34  \\
    \hline
    \specialrule{0em}{0.5pt}{0.5pt}
    \hline
    \end{tabular}%
  \label{tab:addlabel}%
\end{table*}%
\begin{table*}[htbp]
  \centering
  \caption{Experimental Results under the (2 m, HT) Scenario}
    \begin{tabular}{ccccccccccccccc}
    \hline
    \specialrule{0em}{0.5pt}{0.5pt}
    \hline
    \multirow{2}[4]{*}{subject ID} & \multicolumn{2}{c}{True value (bpm)} & \multicolumn{3}{c}{Estimated RR (bpm)} & \multicolumn{3}{c}{Estimated HR (bpm)} & \multicolumn{3}{c}{RR Error rate (\%)} & \multicolumn{3}{c}{HR Error rate (\%)} \\
\cmidrule{2-15}          & Chest Strap RR & Chest Strap HR & Ref1  & Ref2  & Ours  & Ref1  & Ref2  & Ours  & Ref1  & Ref2  & Ours  & Ref1  & Ref2  & Ours \\
    \midrule
    1     & 20.41  & 79.00  & 15.30  & 21.30  & 21.60  & 84.90  & 89.30  & 87.10  & 25.04  & 4.36  & 5.83  & 7.47  & 13.04  & 10.25  \\
    2     & 20.86  & 83.00  & 21.40  & 24.00  & 24.00  & 86.40  & 93.50  & 77.70  & 2.59  & 15.05  & 15.05  & 4.10  & 12.65  & 6.39  \\
    3     & 19.95  & 84.00  & 18.30  & 19.20  & 19.50  & 101.40  & 100.50  & 90.50  & 8.27  & 3.76  & 2.26  & 20.71  & 19.64  & 7.74  \\
    4     & 15.57  & 87.00  & 15.00  & 15.00  & 15.00  & 92.40  & 101.50  & 78.60  & 3.66  & 3.66  & 3.66  & 6.21  & 16.67  & 9.66  \\
    5     & 17.24  & 91.00  & 15.70  & 15.00  & 17.40  & 101.40  & 92.40  & 90.90  & 8.93  & 12.99  & 0.93  & 11.43  & 1.54  & 0.11  \\
    6     & 17.98  & 92.00  & 16.80  & 16.50  & 16.80  & 74.10  & 74.40  & 73.50  & 6.56  & 8.23  & 6.56  & 19.46  & 19.13  & 20.11  \\
    7     & 18.56  & 85.00  & 20.40  & 20.70  & 20.70  & 70.40  & 99.60  & 100.50  & 9.91  & 11.53  & 11.53  & 17.18  & 17.18  & 18.24  \\
    8     & 15.28  & 83.00  & 9.00  & 15.00  & 15.90  & 90.60  & 72.30  & 71.70  & 41.10  & 1.83  & 4.06  & 9.16  & 12.89  & 13.61  \\
    \hline
    \specialrule{0em}{0.5pt}{0.5pt}
    \hline
    \end{tabular}%
  \label{tab:addlabel}%
\end{table*}%

\bibliographystyle{IEEEtran}
\bibliography{ref}

\begin{thebibliography}{10}
\providecommand{\url}[1]{#1}
\csname url@samestyle\endcsname
\providecommand{\newblock}{\relax}
\providecommand{\bibinfo}[2]{#2}
\providecommand{\BIBentrySTDinterwordspacing}{\spaceskip=0pt\relax}
\providecommand{\BIBentryALTinterwordstretchfactor}{4}
\providecommand{\BIBentryALTinterwordspacing}{\spaceskip=\fontdimen2\font plus
\BIBentryALTinterwordstretchfactor\fontdimen3\font minus \fontdimen4\font\relax}
\providecommand{\BIBforeignlanguage}[2]{{%
\expandafter\ifx\csname l@#1\endcsname\relax
\typeout{** WARNING: IEEEtran.bst: No hyphenation pattern has been}%
\typeout{** loaded for the language `#1'. Using the pattern for}%
\typeout{** the default language instead.}%
\else
\language=\csname l@#1\endcsname
\fi
#2}}
\providecommand{\BIBdecl}{\relax}
\BIBdecl

\bibitem{un_wpp2022}
\BIBentryALTinterwordspacing
{United Nations, Department of Economic and Social Affairs, Population Division}. (2022) World population prospects 2022. [Online]. Available: \url{https://www.un.org/development/desa/pd/sites/www.un.org.development.desa.pd/files/undesa_pd_2022_wpp-data_sources.pdf}
\BIBentrySTDinterwordspacing

\bibitem{2b}
H.~Wu, Y.~Wang, H.~Zhang, X.~Yin, L.~Wang, L.~Wang, and J.~Wu, ``An investigation into the health status of the elderly population in china and the obstacles to achieving healthy aging,'' \emph{Scientific Reports}, vol.~14, no.~1, p. 31123, 2024.

\bibitem{4}
F.~K. Winston and A.~C. Yan, ``Wearable health device dermatitis: a case of acrylate-related contact allergy,'' \emph{Cutis}, vol. 100, no.~2, pp. 97--99, 2017.

\bibitem{3}
G.-J. Jong, Aripriharta, and G.-J. Horng, ``The ppg physiological signal for heart rate variability analysis,'' \emph{Wireless Personal Communications}, vol.~97, no.~4, pp. 5229--5276, 2017.

\bibitem{7}
Y.~Zhang, F.~Qi, H.~Lv, F.~Liang, and J.~Wang, ``Bioradar technology: Recent research and advancements,'' \emph{IEEE Microwave Magazine}, vol.~20, no.~8, pp. 58--73, 2019.

\bibitem{8}
C.~Li, J.~Cummings, J.~Lam, E.~Graves, and W.~Wu, ``Radar remote monitoring of vital signs,'' \emph{IEEE Microwave magazine}, vol.~10, no.~1, pp. 47--56, 2009.

\bibitem{9}
A.~Ahmad, J.~C. Roh, D.~Wang, and A.~Dubey, ``Vital signs monitoring of multiple people using a fmcw millimeter-wave sensor,'' in \emph{2018 IEEE Radar Conference (RadarConf18)}.\hskip 1em plus 0.5em minus 0.4em\relax IEEE, 2018, pp. 1450--1455.

\bibitem{10}
M.~Alizadeh, G.~Shaker, J.~C.~M. De~Almeida, P.~P. Morita, and S.~Safavi-Naeini, ``Remote monitoring of human vital signs using mm-wave fmcw radar,'' \emph{IEEE Access}, vol.~7, pp. 54\,958--54\,968, 2019.

\bibitem{11}
H.~Zhao, H.~Hong, D.~Miao, Y.~Li, H.~Zhang, Y.~Zhang, C.~Li, and X.~Zhu, ``A noncontact breathing disorder recognition system using 2.4-ghz digital-if doppler radar,'' \emph{IEEE journal of biomedical and health informatics}, vol.~23, no.~1, pp. 208--217, 2018.

\bibitem{12}
L.~Wen, S.~Dong, Y.~Wang, C.~Gu, Z.~Tang, Z.~Liu, Y.~Wang, and J.~Mao, ``Noncontact infant apnea detection for hypoxia prevention with a k-band biomedical radar,'' \emph{IEEE Transactions on Biomedical Engineering}, vol.~71, no.~3, pp. 1022--1032, 2023.

\bibitem{13}
C.~Gu, R.~Li, H.~Zhang, A.~Y. Fung, C.~Torres, S.~B. Jiang, and C.~Li, ``Accurate respiration measurement using dc-coupled continuous-wave radar sensor for motion-adaptive cancer radiotherapy,'' \emph{IEEE Transactions on biomedical engineering}, vol.~59, no.~11, pp. 3117--3123, 2012.

\bibitem{14}
L.~Ren, H.~Wang, K.~Naishadham, O.~Kilic, and A.~E. Fathy, ``Phase-based methods for heart rate detection using uwb impulse doppler radar,'' \emph{IEEE Transactions on Microwave Theory and Techniques}, vol.~64, no.~10, pp. 3319--3331, 2016.

\bibitem{15}
W.~Ren, F.~Qi, F.~Foroughian, T.~Kvelashvili, Q.~Liu, O.~Kilic, T.~Long, and A.~E. Fathy, ``Vital sign detection in any orientation using a distributed radar network via modified independent component analysis,'' \emph{IEEE Transactions on Microwave Theory and Techniques}, vol.~69, no.~11, pp. 4774--4790, 2021.

\bibitem{17}
Q.~Liu, H.~Guo, J.~Xu, H.~Wang, A.~Kageza, S.~AlQarni, and S.~Wu, ``Non-contact non-invasive heart and respiration rates monitoring with mimo radar sensing,'' in \emph{2018 IEEE Global Communications Conference (GLOBECOM)}.\hskip 1em plus 0.5em minus 0.4em\relax IEEE, 2018, pp. 1--6.

\bibitem{19}
T.~K.~V. Dai, K.~Oleksak, T.~Kvelashvili, F.~Foroughian, C.~Bauder, P.~Theilmann, A.~E. Fathy, and O.~Kilic, ``Enhancement of remote vital sign monitoring detection accuracy using multiple-input multiple-output 77 ghz fmcw radar,'' \emph{IEEE Journal of Electromagnetics, RF and Microwaves in Medicine and Biology}, vol.~6, no.~1, pp. 111--122, 2021.

\bibitem{20}
B.~Zhang, B.~Jiang, R.~Zheng, X.~Zhang, J.~Li, and Q.~Xu, ``Pi-vimo: Physiology-inspired robust vital sign monitoring using mmwave radars,'' \emph{ACM Transactions on Internet of Things}, vol.~4, no.~2, pp. 1--27, 2023.

\bibitem{21}
W.~Ren, J.~Cao, H.~Yi, K.~Hou, M.~Hu, J.~Wang, and F.~Qi, ``Noncontact multipoint vital sign monitoring with mmwave mimo radar,'' \emph{IEEE Transactions on Microwave Theory and Techniques}, 2024.

\bibitem{22}
S.~Wang, C.~Han, J.~Guo, and L.~Sun, ``Mm-fgrm: Fine-grained respiratory monitoring using mimo millimeter wave radar,'' \emph{IEEE Transactions on Instrumentation and Measurement}, vol.~73, pp. 1--13, 2023.

\bibitem{23}
G.~Shafiq and K.~C. Veluvolu, ``Multimodal chest surface motion data for respiratory and cardiovascular monitoring applications,'' \emph{Scientific data}, vol.~4, no.~1, pp. 1--12, 2017.

\bibitem{24}
A.~Singh, S.~U. Rehman, S.~Yongchareon, and P.~H.~J. Chong, ``Modelling of chest wall motion for cardiorespiratory activity for radar-based ncvs systems,'' \emph{Sensors}, vol.~20, no.~18, p. 5094, 2020.

\end{thebibliography}

\begin{IEEEbiography}[{\includegraphics[width=1in,height=1.25in,clip,keepaspectratio]{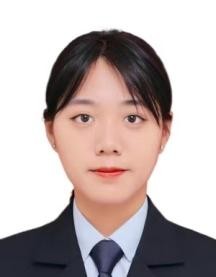}}]{Heyao Zhu} received the Bachelor of Science degree in biomedical engineering from Fourth Military Medical University, Xi'an, China, in 2026 (expected). She was born in Baoding, Hebei Province, China, in 2003. \\
\indent Since 2022, she has been pursuing her undergraduate studies at Fourth Military Medical University, where her research interests include biomedical signal processing and medical device innovation. She received the Second Prize in the National Biomedical Engineering Innovation Design Competition (2024), the First Prize (Shaanxi Division) in the Higher Education Press Cup National Mathematical Contest in Modeling (2024), and the Second Prize in the Military Mathematical Modeling Contest of PLA (2024).
\end{IEEEbiography}

\begin{IEEEbiography}[{\includegraphics[width=1in,height=1.25in,clip,keepaspectratio]{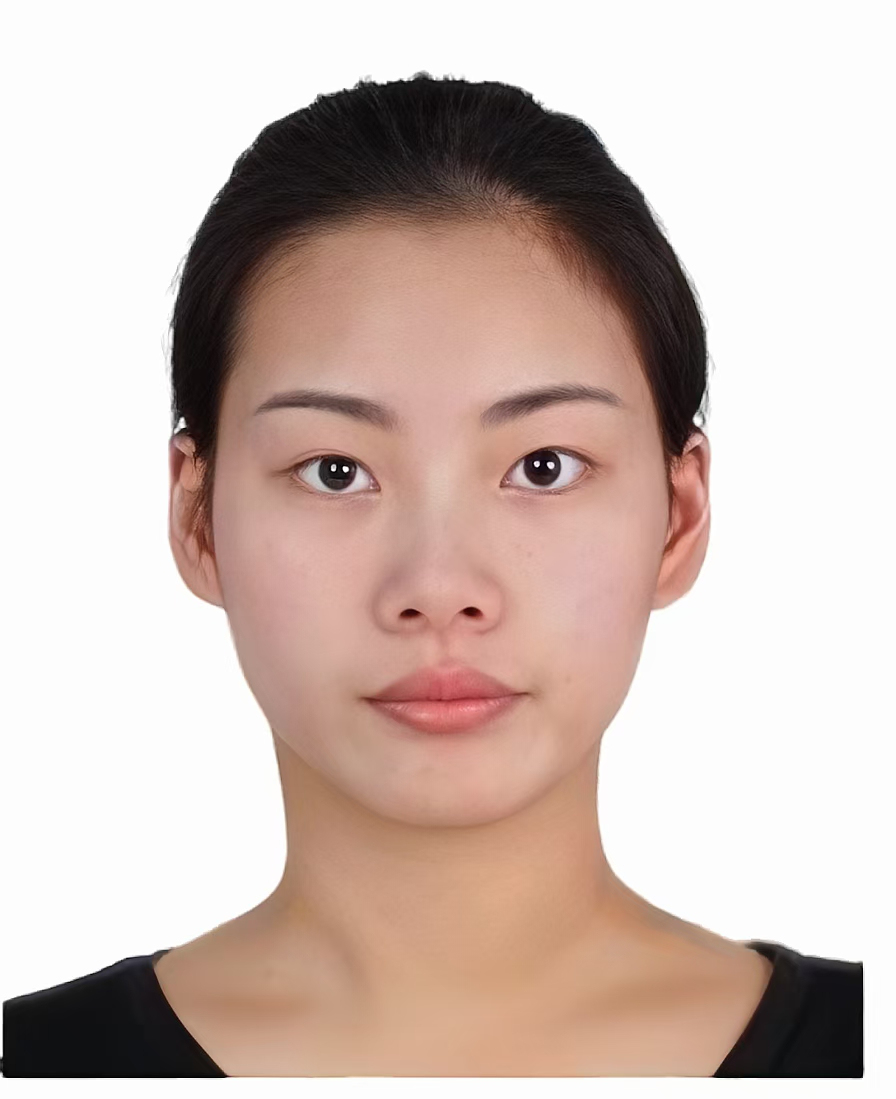}}]{Yimeng Zhao}is currently working toward the M.S. degree with the School of Mechanical and Electrical Engineering, Chengdu University of Technology. Her research interests include Bio-radar-based through-wall human activity recognition and non-contact intelligent life sensing. 
\end{IEEEbiography}

\begin{IEEEbiography}[{\includegraphics[width=1in,height=1.25in,clip,keepaspectratio]{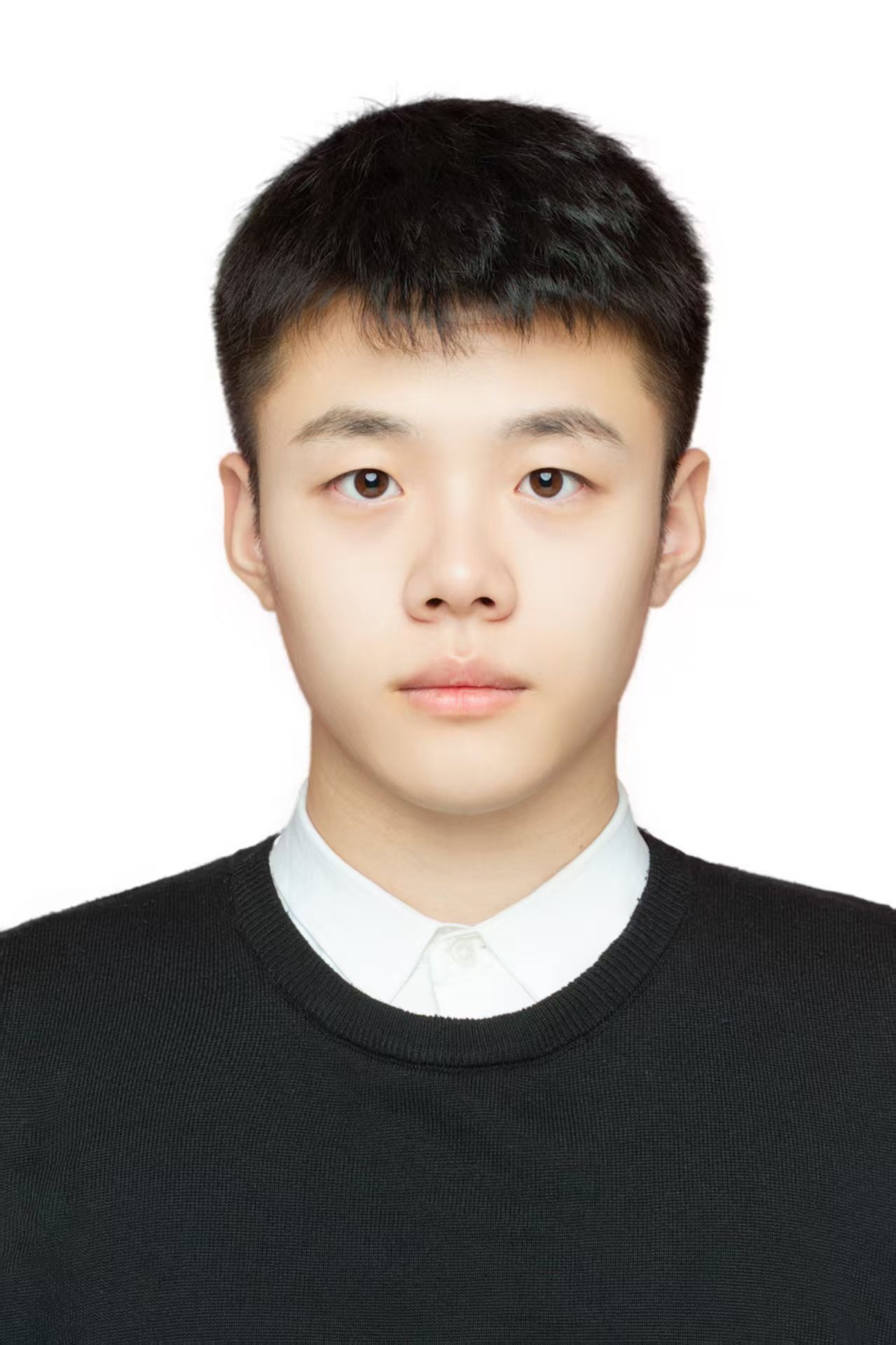}}]{Zirui Zhang} received the Bachelor of Science degree in biomedical engineering from Air Force Medical University (Fourth Military Medical University), Xi'an, China, in 2027 (expected). \\
\indent Since 2023, he has been conducting research under the guidance of Researcher QI Fugui at the Teaching and Experiment Center, Department of Military Biomedical Engineering. His work focuses on non-contact identity recognition using MIMO-FMCW bio-radar technology for cardiac micro-motion sensing. His research interests include precise detection of physiological behavior characteristics and continuous biometric identification. He has authored one software copyright and received three provincial-level awards and three national awards in disciplinary competitions.
\end{IEEEbiography}

\begin{IEEEbiography}[{\includegraphics[width=1in,height=1.25in,clip,keepaspectratio]{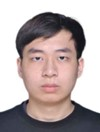}}]{Huansheng Yi} received the B.E. degrees in internet of things engineering from LuoYang Institute of Science and Technology in 2023, Henan, China. \\
\indent He is currently serving as a RA in the BME department, FMMU, Xi'an. Since 2019, He majored in internet of things and embedded technology, school of computer and information engineering, Luoyang, Henan, China. He won the third prize in the ICAN Innovation Contest(final), 2021. He won the second prize of China Undergraduate Computer Design Competition, 2022. He won the third prize of China Undergraduate Service Outsourcing Innovation and Entrepreneurship Competition, 2022. He also awarded scholarship for many times in college.
\end{IEEEbiography}

\begin{IEEEbiography}[{\includegraphics[width=1in,height=1.25in,clip,keepaspectratio]{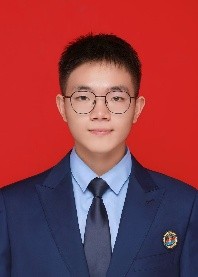}}]{Chenbin Gao} was born in Xiayi County, Shangqiu, Henan Province, China, in 2006. He entered Fourth Military Medical University, Xi'an, China, in 2024, where he is currently a first-year cadet pursuing the Bachelor of Science degree in biomedical engineering. \\
\indent He is a student researcher at the Non-Contact Intelligent Sensing Laboratory, focusing on non-contact physiological monitoring technologies. He was awarded the Second Prize in the university-level competition of the College Student Innovation and Entrepreneurship Contest in 2024.
\end{IEEEbiography}

\begin{IEEEbiography}[{\includegraphics[width=1in,height=1.25in,clip,keepaspectratio]{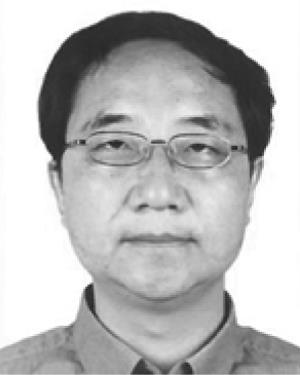}}]{Jianqi Wang} was born in Xi'an, Shaanxi, China, in 1962. He received the B.E. degree in information and control engineering from Xi'an Jiaotong University, Xi'an, in 1984; the M.E. degree in electronic engineering from the National University of Defense Technology, Changsha, China, in 1990; and the Ph.D. degree from the Key Laboratory of the Ministry of Education of China, Xi'an Jiaotong University, in 2006. \\
\indent Since 1990, he has taught in the School of Biomedical Engineering, Fourth Military Medical University, Xi'an. He is currently serving as a Professor with and the Director of the Department of Electronics. He pioneered radar-based human being detection in China in 1998 and has published more than 100 articles on the technology. His research interest is bio-radar technology, including signal processing, human being detection, and imaging.
\end{IEEEbiography}
    
\begin{IEEEbiography}[{\includegraphics[width=1in,height=1.25in,clip,keepaspectratio]{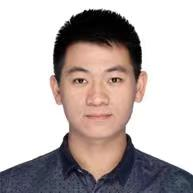}}]{Fugui Qi} (Member, IEEE)  received the B.E. and Ph.D. degrees in biomedical engineering from the Institute of Biomedical Engineering (BME), Fourth Military Medical University (FMMU), Xi'an, China, in 2014 and 2020, respectively. \\
\indent Since 2020, he has been with the School of Biomedical Engineering, Fourth Military Medical University, Xi'an.  He is currently serving as a research fellow in the Bio-radar and Signal Processing Laboratory, School of BME, FMMU.  From 2019 to 2020, he was a visiting scholar with The University of Tennessee, Knoxville, TN, USA.  He has published more than 40 peer-reviewed journal and conference papers.  His research interests include bio-radar-based medical signal processing, through-wall human motion recognition, and non-contact intelligent life sensing. He received the Excellent Doctoral Dissertations of Shaanxi Province in 2022 and was selected for the Young Talent Fund of the Association for Science and Technology in Shaanxi, China, in 2023. He also won the First Prize for Young Excellent Paper Competition (Young Scholar) of the 2023 China BME conference and the Second Award for Science Research Famous Achievement Award in Shaanxi Higher Institution in 2024. 
\end{IEEEbiography}

\vfill

\end{document}